# A Chemistry-First Centered Icy Chemical Inventory of Protostellar Sources with JWST

Andrew M. Turner,[1†] Yao-Lun Yang,[2†*] Rachel Gross,[3] Nami Sakai,[2] Ralf I. Kaiser[1*]
[1]Department of Chemistry & W.M. Keck Research Laboratory in Astrochemistry, University of Hawaii at Manoa, Honolulu, HI, 96822, USA
[2]Star and Planet Laboratory, RIKEN Pioneering Research Institute, Wako, Saitama, Japan
[3]Department of Astronomy, University of Virginia, Charlottesville VA 22904
[†]Authors contributed equally

**Abstract**

The chemical evolution in star forming regions is driven by the interplay between gas and ice mantles. Identifying the ice compositions at the early stage of star formation thus provides constraints on the chemical processes inaccessible from gas-phase characterizations. As part of the CORINOS program, spectra from the James Webb Space Telescope (JWST) MIRI MRS were taken toward four Class 0 protostars: IRAS 15398-3359, Ser-emb7, L483, and B335. The spectra were processed with simultaneous fitting of a continuum and silicate absorption to produce optical depth mid-infrared spectra of the ices at 5-28 μm (360 – 2000 cm$^{-1}$) toward these four sources. Simple molecules such as water ($H_2O$), carbon dioxide ($CO_2$), methanol ($CH_3OH$), formic acid/formate ($HCOOH$/$HCOO^-$), ammonia/ammonium ($NH_3$/$NH_4^+$), and formaldehyde ($H_2CO$) are the most abundant features in these ices, while complex organic molecules (COMs) represent a smaller contribution. Likely COMs include hydroxylamine ($NH_2OH$), methylamine ($CH_3NH_2$), and ethanol ($CH_3CH_2OH$). Absorption features belonging to functional groups such as $–CH_3$ and –OH suggest that additional COMs are present, but these cannot be unambiguously assigned due to overlapping bands. Formation pathways toward these COMs utilizing radical-radical combination reactions based on laboratory simulation experiments is presented. By extension, COMs predicted by these reactions, but absent from the spectra, are discussed. The results provide insight into the chemical environment of these ices and also highlight the critical need for caution and sufficient evidence in order to confidently identify COMs in ice.

## 1. Introduction

The chemical evolution during star formation determines the initial condition of planet-forming disks. In recent years, complex organic molecules (COMs) have been detected in increasing numbers of sources and in environments ranging from prestellar cores to protostars and protoplanetary disks (Belloche et al. 2020; Booth et al. 2021; Scibelli et al. 2024; Yang et al. 2021). While the formation of these COMs is considered to occur in the ices of ice coated dust grains, most observational constraints come from gas-phase characterization of COMs, which is regulated by desorption processes and could be altered by gas-phase reactions after the desorption of COMs (Codella et al. 2017; Garrod & Widicus Weaver 2013; Jørgensen et al. 2020; López-Sepulcre et al. 2024). Thus, a direct spectroscopic identification of COMs in the ice mantle becomes critical evidence to test our understanding of gas-grain chemistry.

In space, the composition of ices is typically probed by infrared spectroscopy. The vibrational energy of specific molecular structures leads to absorption features in spectra as the background radiation penetrates icy grains, which are used for identifying ice species in observations. While simple species such as water ($H_2O$), carbon monoxide (CO), and carbon dioxide ($CO_2$) have isolated absorption features (Boogert et al. 2015), icy COMs have been challenging to identify observationally (Abplanalp et al. 2016a; Turner & Kaiser 2020). First, the sensitivity of astronomical observations using ISO, AKARI, and Spitzer at mid-infrared wavelengths was insufficient to robustly characterize weaker absorption features other than the isolated ones (Aikawa et al. 2012; Boogert et al. 2008; Bottinelli et al. 2010; Gibb et al. 2004; Kim et al. 2022; Öberg et al. 2008; Pontoppidan et al. 2008), including $H_2O$, CO, $CO_2$, and methane ($CH_4$). Faint absorption features were discussed, but robust identifications were difficult to conclude (Schutte et al. 1999; Schutte & Khanna 2003). Secondly, fewer laboratory measurements of complex ice species at infrared wavelengths existed before the realization of the James Webb Space Telescope (JWST) (Bennett & Kaiser 2007a, 2007b; Bergner et al. 2016; Bisschop et al. 2007; Ennis et al. 2017; Galvez et al. 2010; Hudson & Gerakines 2018; Shimanouchi 1973; Yeo & Ford 1990; Zheng & Kaiser 2010). The substantially improved sensitivity of JWST motivated a new wave of laboratory studies on complex ice species and their mixtures (Rachid et al. 2021; Rachid et al. 2020; Rachid et al. 2022; Santos et al. 2024; Slavicinska et al. 2025; Slavicinska et al. 2023; van Scheltinga et al. 2018; van Scheltinga et al. 2021; Zhang et al. 2024; Zhang et al. 2025a, 2025b; Zhang et al. 2025c). Lastly, complex molecular species often have similar vibrational bands due to the similarity in the functional groups—a group of atoms with a characteristic chemical reactivity—resulting in overlapping absorption features (Abplanalp et al. 2016a; Fleming & Williams 2019; Socrates 2001; Turner & Kaiser 2020). Thus, isolating the unique absorption features of specific compounds and quantifying their abundance remains difficult and often questionable.

With JWST, high signal-to-noise (S/N) infrared spectra are obtainable toward molecular clouds and protostars, probing the ice compositions in those environments (Federman et al. 2024; McClure et al. 2023; Tyagi et al. 2025; Yang et al. 2022). Several approaches have been developed to model the simple ice species as well as identifying icy COMs. For isolated absorption features, a local baseline is often fitted to the absorption-free spectrum for deriving the optical depth of the features (Chu et al. 2020; McClure et al. 2023). This method produces a simple optical depth spectrum that can be directly modeled using laboratory measurements, deriving the ice column densities. When an isolated but complex absorption feature is extracted using a local baseline, multiple components motivated by laboratory studies can be tested, thus resulting in identification and quantification of ice abundance (Brunken et al. 2025; Brunken et al. 2024; Nazari et al. 2024). Another method is to model the entire spectrum that includes several ice features using statistical methods. Rocha et al. (2021) present a genetic modeling algorithm (ENIIGMA) that searches for the best combination of ices from a database such as the Leiden Ice Database for Astrochemistry (LIDA)(Rocha et al. 2022), and optimizes the weights of the chosen laboratory spectra to derive the ice column densities (Chen et al. 2024; Rocha et al. 2025; Rocha et al. 2024). Rayalacheruvu et al. (2025) demonstrates a global minimization framework to model the entire ice spectra and infer the presence of complex ice species.

The local baseline method leaves the identification of ice species to manual trials, while the statistical method aims to optimize the choice of ice composition in addition to their column densities. Most icy COMs have absorption features at 7–8 μm (1430–1250 $cm^{-1}$) that are often overlapping due to their common functional groups, making this spectral range critical for COM identification. However, identifying COM species from a blended composite absorption feature has a greater uncertainty and can often lead to speculative assignments. Statistical methods, such as the ENIIGMA, can find an ice composition that best describes the observed optical depth spectrum, but it lacks chemical knowledge to determine if the resulting combination is astrochemically feasible, which requires human intervention. Another way to improve the identification is looking for fainter but isolated absorption features of COMs in ices. For example, ethanol has its C–C vibration mode, ν(CC), at 11.2 μm (890 $cm^{-1}$), which is well separated from the features of other COMs. In this study, we aim to develop an identification scheme driven by laboratory measurements as well as astrochemical knowledge of star-forming regions to find likely contributors of observed absorption features in JWST mid-infrared spectra.

The observational data are taken from the CORINOS program (Yang et al. 2022), which consists of JWST MIRI MRS observations of four isolated Class 0 protostars. These four sources were selected to study the ice composition of the protostars with rich/poor gas-phase COMs as well as different luminosities. The protostars with rich gas-phase COMs, B335 and L483, and the ones with poor gas-phase COMs, IRAS 15398-3359 and Ser-emb 7, were selected based on the ALMA observations available at the time of observation design (Bergner et al. 2019; Imai et al. 2016; Jacobsen et al. 2019; Okoda et al. 2020). Recently, Okoda et al. (2023) found weak emission of methyl formate in IRAS 15398-3359, but gas-phase detection of other COMs remains elusive. These four sources were also selected to have bolometric luminosities of ~1 $L_{\odot}$ and ~10 $L_{\odot}$, which serve as a proxy of thermal desorption to explore the ice chemistry at different environments. This study presents a chemistry-first driven identification of ice species in the CORINOS protostars. Section 2 describes the details of the observations as well as the method to derive ice optical depth spectra. Section 3 presents the identification scheme and the results of ice identification. Section 4 discuss the chemical relation and reactions that could contribute to the identified species. Finally, Section 5 summarizes the conclusions of this study.

## 2. Observations & Methods

### 2.1. JWST observations

The spectroscopic data were taken by the CORINOS program (Yang et al. 2022). The 4.9–28 μm (2040–360 $cm^{-1}$) spectra of four protostars were observed by the Mid-Infrared Instruments (MIRI) Medium Resolution Spectroscopy (MRS) on JWST (Gardner et al. 2023; Rieke et al. 2015; Wright et al. 2023). Table 1 shows the observation date along with basic properties of the sample. MIRI MRS has four integral field units (IFU), each of which has three wavelength ranges that observe separately. The integration time is 1433, 3631, and 1433 seconds for the short, medium, and long wavelength ranges, respectively. The medium range was integrated for longer to achieve a better S/N on the strong silicate absorption feature at 10 μm (1000 $cm^{-1}$). We used a 4 point dither pattern optimized for extended source, a technique of observing the same field of view with small offsets to mitigate the undersampled point spread function, and the detector readout pattern is SLOWR1. The specific observations analyzed can be accessed from the Mikulski Archive for Space Telescopes (MAST) at the Space Telescope Science Institute via doi: 10.17909/77j7-pd29.

The spectroscopic data were reduced using the JWST Calibration pipeline v.1.15.1 (Bushouse 2024) with calibration reference data `jwst_1256.pmap` from the raw Stage 0 data to Stage 3 calibrated data. We enabled the pixel-by-pixel background subtraction in Stage 2, where the slope data for measuring flux density were calibrated using dedicated background observations. We also enabled the `fringe` and `residual_fringe` steps in Stage 2 to reduce the fringe using reference files as well as algorithmically determined fringe patterns.

**Table 1**.Source properties and observation details

| Source | RA (ICRS) | Dec (ICRS) | $L_{bol}$ ($L_{\odot}$) | Obs. date | $M_{env}$ ($M_{\odot}$) | $L_{bol}$, $M_{env}$ Reference |
|---|---|---|---|---|---|---|
| IRAS 15398-3359 | 15:43:02.24 | -34:09:06.83 | 1.4 | 2022 July 20 | 1.2 | 1, 2 |
| B335 | 19:37:00.90 | +07:34:09.49 | 1.4 | 2023 May 14 | 3.3 | 1, 3 |
| Ser-emb 7 | 18:28:54.06 | +00:29:29.29 | 7.9 | 2023 May 7 | 1.67 | 4 |
| L483 | 18:17:29.94 | -04:39:39.62 | 10.5 | 2023 April 18 | 4.4 | 5 |

**Notes**: The source positions are measured from the ALMA continuum peak using the measurements in ALMA programs 2021.1.00357.S, 2019.1.01792.S, and 2017.1.00693.S for IRAS 15398-3359, Ser-emb 7, and L483, respectively. For B335, the continuum position is taken from (Okoda et al. 2022).
**References**. (1) (Ohashi et al. 2023); (2) (Jørgensen et al. 2013); (3) (Evans et al. 2023); (4) (Enoch et al. 2009); (5) (Jacobsen et al. 2019)

## 2.2. Spectral Extraction

The spectra analyzed in this study were extracted from the IFU cubes following a similar approach outlined in Yang et al. (2022). We use an aperture centered on the submillimeter continuum peak position measured by ALMA (Table 1), whose size is set to four times of the beam size, following the relation derived in Law et al. (2023). At each wavelength channel, we measured the flux density within the aperture using `aperture_photometry` task from the `photutils` package v.1.12.0 (Bradley et al. 2024). There are 12 calibrated IFU cubes (4 IFUs each with 3 wavelength ranges), covering the full wavelength coverage of MIRI; thus, we obtained 12 spectra for each source. To combine these 12 spectra, we scaled the spectra from the shortest wavelength to match the median values of flux density in the overlapped wavelength ranges. Compared to the sample, B335 has significantly weaker flux density toward the protostar, resulting in extremely noisy spectra. In fact, a point source is only detected at $\gtrsim$13 μm. Thus, for B335, we extracted the spectra at 1” offset in RA from the continuum position, which is toward the less extinct, blue-shifted outflow cavity. We also used a fixed aperture of 1” to maximize the sensitivity at the shorter wavelengths. There is no scaling between spectral segment for B335. The extracted spectra are shown in Figure 1.

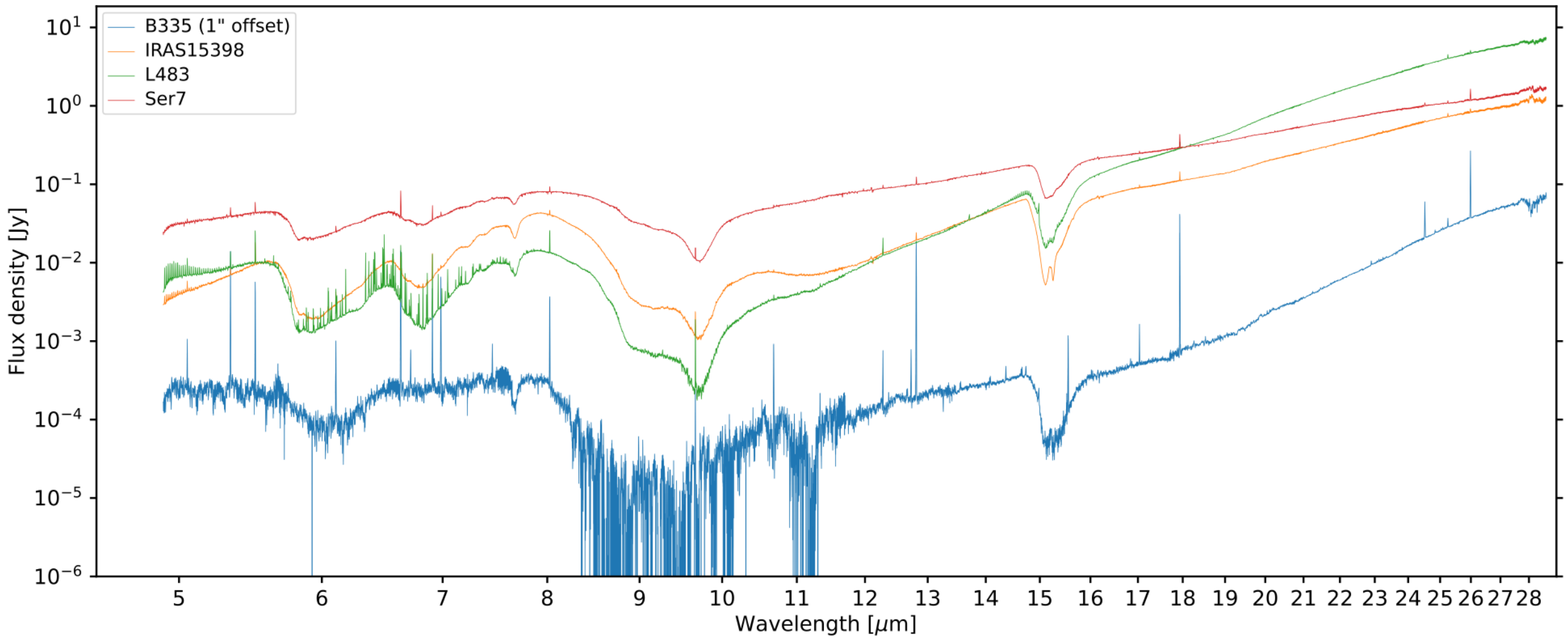

**Figure 1**. MIRI/MRS spectra of the four CORINOS sources. The B335 spectrum is extracted from 1" offset position in the blue-shifted outflow, while the other spectra are extracted on-source with an aperture of four times the beam size.

2.3. Optical Depth Spectra

The MRS spectra show several absorption features that indicate the presence of abundant ice, likely in the protostellar envelopes. In principle, we can use extinction-free continuum to derive the optical depth spectrum and characterize the ice composition. However, the strong absorption of silicate dust prevails in the spectra at 10-25 μm, making a straightforward determination of the baseline challenging. Thus, removing the extinction-free continuum together with the absorption of silicate dust is the necessary step before ice analysis. Ice studies using pre-JWST spectra often use template spectra, such as the GCS 3 (Kemper et al. 2004) or ice-free spectra, to subtract the dust absorption (Boogert et al. 2008; Bottinelli et al. 2010). For JWST spectra, the GCS 3 spectrum has little success in reproducing the observed silicate feature, whereas a mixture of synthetic silicate dust works relatively well (Rocha et al. 2025). In this study, we fitted a polynomial baseline together with synthetic pyroxene and olivine simultaneously to derive the ice optical depth spectra. The baseline is a 4th-order polynomial described in logarithmic scale, $F_{poly} = f(\log(\lambda))$. The dust opacities are simulated using `optool` (Dominik et al. 2021) using the experimental data from Dorschner et al. (1995). Figure 2 shows the fitting results as well as the wavelength ranges chosen for the fitting. Typically, we weight the data of each wavelength equally. But for L483, the fitted composite baseline was too flat at about 5 μm with an equal weighting; thus, we first normalized the weight of each anchor wavelength range (red) so that each range is weighted equally. Then, we increased the weight for the 5 μm range to be 30 times higher than other ranges to achieve a better fitting result. We also applied a median filter with a kernel size of 31 channels for the L483 spectrum to trace a spectrum without emission lines. After the fitting, we derived the optical depth spectra using $-\ln(F_\lambda/F_{poly}\, e^{(-(\tau_{pyr} + \tau_{oli}))})$, where $\tau_{pyr}$ and $\tau_{oli}$ are the fitted optical depth of pyroxene and olivine Figure 3 shows the ice optical depth spectra.

2.4. Spectral Analysis

These spectra were fit with a combination of Gaussian peaks using the Grams/AI program (2002) until the residual (i.e., the difference between the spectrum and the sum of the fitted peaks) reached a minimum value consistent with the noise level of the spectrum. This was accomplished by iteratively adjusting peak model parameters using the Levenberg-Marquardt (LMFit) algorithm to minimize the difference ($\chi^2$) between a calculated spectrum of the fitted peaks and the original experimental data after a local baseline was fitted to flatten the optical depth spectra. An analysis of the spectra of the four CORINOS sources reveals similar molecular signatures, suggesting that key molecules are pervasive in interstellar ices around protostars. However, the spectra are not identical and display a range of complexity from simple starting materials to first- and second-generation complex organic molecules. Molecular assignments were achieved by referencing infrared spectra of laboratory ices, and the assignments for each spectrum along with their spectral references are compiled in Table 2. In an ideal case, the assignment of a molecule would utilize multiple unique absorption peaks that do not overlap with other molecules. In reality, astronomical spectra of mixed-composition ices are far from ideal, and thus a few principles guide our analysis.

1. Multiple peaks increase the confidence of assignments; however, exceptions exist for small molecules with a single well-known transition, such as carbon dioxide ($CO_2$, $\nu_2$ at 660 cm$^{-1}$), methane ($CH_4$, $\nu_4$ at 1300 cm$^{-1}$), and carbon monoxide (CO, $\nu_1$ at 2135 cm$^{-1}$), which is outside the range of these spectra.
2. The observed absorption bands for a molecule should have approximately the same relative band strengths as the reference spectra. As a corollary, the weakest bands should not be used to identify a compound if the strongest bands are absent.
3. When the absorption belongs to a functional group of a homologous series (e.g., an absorption relating to the hydroxyl (–OH) functional group of organic alcohols), the simplest homologue of the series is the preferred choice. If, however, the functional group is present across several classes of compounds, such as an absorption relating to a $-CH_3$ moiety found in alkanes, alcohols, aldehydes, carboxylic acids then the assignment is ambiguous and can only be assigned to the functional group.

Following this identification framework, the spectral analysis for IRAS 15398-3359, Ser-emb7, and L483 are presented below. Due to the high level of noise, a full analysis of B335 was not feasible, and the partially evaluated spectrum is presented in the Appendix.

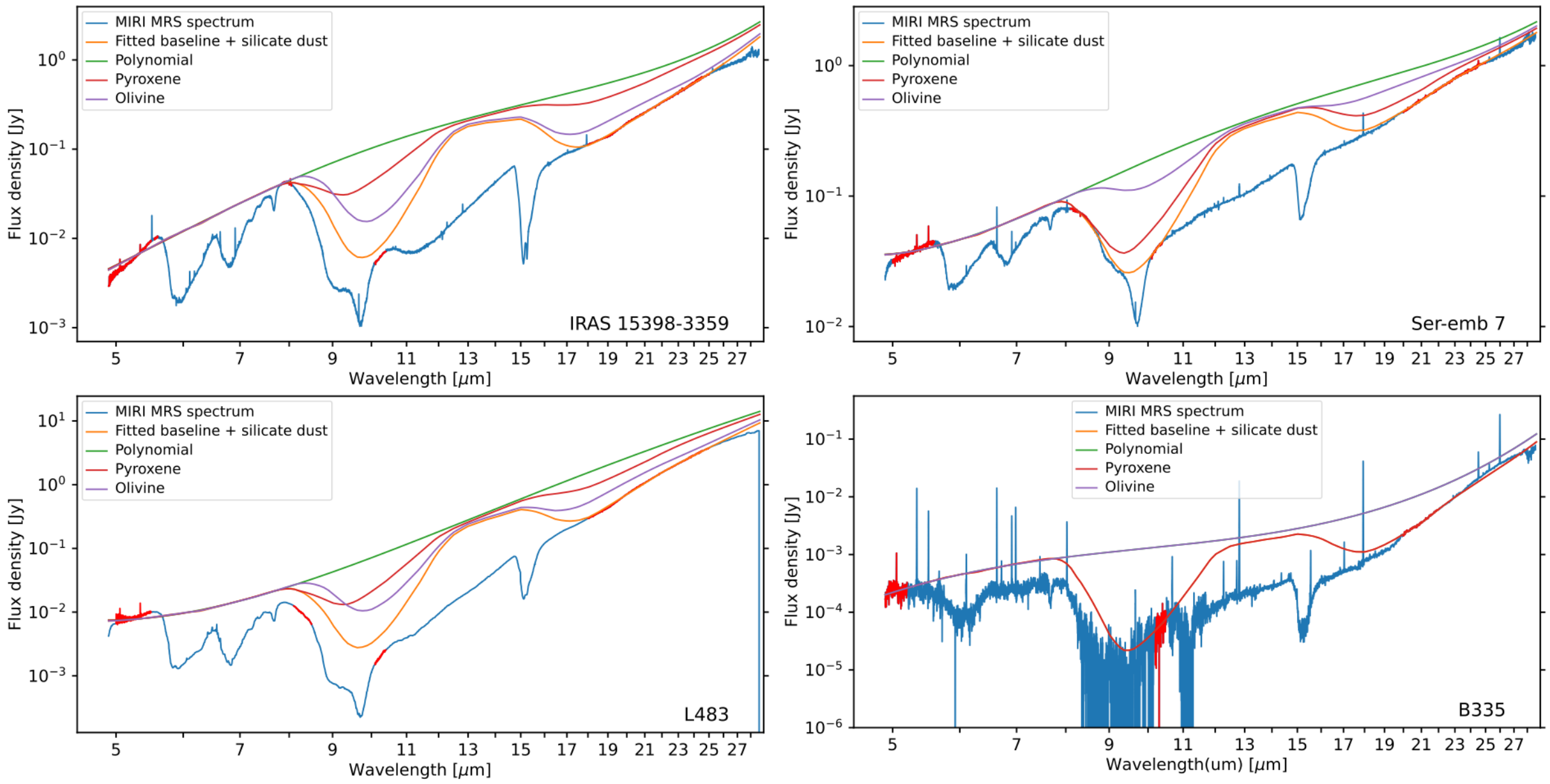

**Figure 2.** The MIRI MRS spectra of the four CORINOS sources along with the best-fitting polynomial baseline (green) and the synthetic silicate dust models of pyroxene (red) and olivine (purple). The composite baseline model is shown in orange. The red spectra highlight the wavelength points chosen for fitting of baseline/silicate.

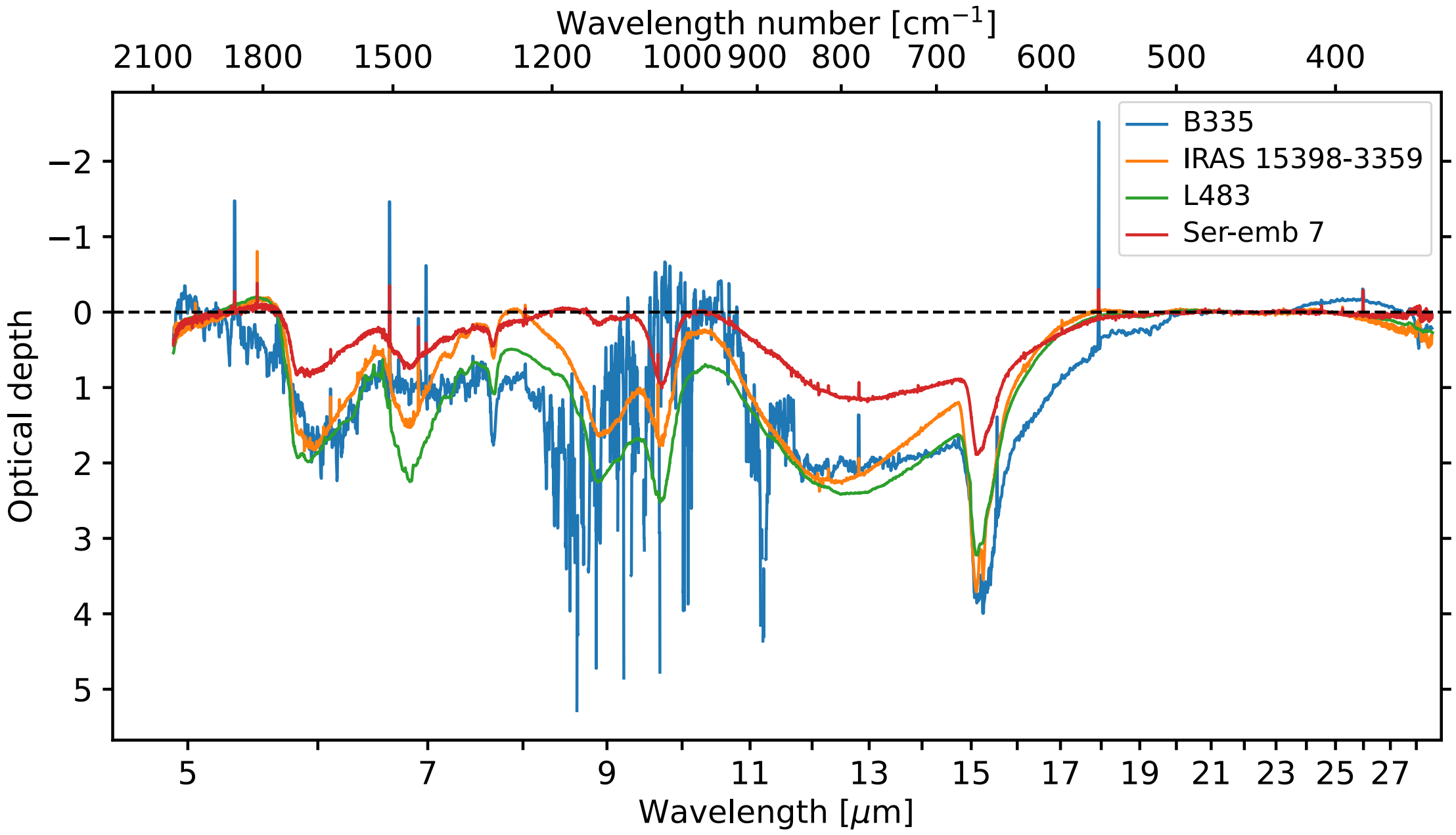

**Figure 3**. The derived optical depth spectra of the CORINOS sources after the removal of silicate absorption features.

## 3. Results

### 3.1. IRAS 15398-3359

The spectrum of IRAS 15398-3359 (Figure 4) includes compounds well-known in interstellar ices (Boogert et al. 2015; McClure et al. 2023). Water ($H_2O$) represents a dominant species with the bending mode ($\nu_2$) and libration mode ($\nu_L$) each spanning several hundred wavenumbers but peaking at 1675 cm$^{-1}$ and 820 cm$^{-1}$, respectively. These broad bands overlap other species, the most prominent of which is carbon dioxide ($CO_2$) centered at 660 cm$^{-1}$ ($\nu_2$). While multiple bands are preferred in order to verify the assignment for a particular molecule, the intense carbon dioxide $\nu_2$ vibration is sufficient for confirmation. Moreover, the double-peaked feature in the carbon dioxide $\nu_2$ mode is a strong indication of pure carbon dioxide (Ehrenfreund et al. 1997). In addition, the weaker but still prominent $\nu_4$ mode of methane ($CH_4$) at 1302 cm$^{-1}$, the only methane transition in this spectral region, allows for a confident assignment. Several bands of methanol ($CH_3OH$) are present, with the most intense peak belonging to the carbon-oxygen stretch at 1029 cm$^{-1}$ ($\nu_8$). Nearby at 1130 cm$^{-1}$ is another band assigned only to methanol ($\nu_{11}$). A cluster of three weaker bands all associated with the deformation of methanol's $CH_3$ functional group may be present ($\nu_5$ at 1447 cm$^{-1}$, $\nu_{10}$ at 1463 cm$^{-1}$, and $\nu_4$ at 1482 cm$^{-1}$), but these are not assigned unambiguously and may be associated with other compounds: see Section 3.4 for further discussion. The bulk of nitrogen is incorporated into ammonia ($NH_3$) and its ionic form, ammonium ($NH_4^+$). Two separate deformation modes of ammonia are observed: the $\nu_2$ mode at 1111 cm$^{-1}$ and the $\nu_4$ band at 1623 cm$^{-1}$, although the latter peak also includes contribution from the $\nu_2$ mode of water. The intense complex of $NH_4^+$ features, peaking between 1410 and 1500 cm$^{-1}$ ($\nu_2$), dwarfs the weaker ammonia band at 1111 cm$^{-1}$. An observed counterion to $NH_4^+$ is the formate ion ($HCOO^-$), and three peaks are identified including the symmetric ($\nu_2$ at 1353 cm$^{-1}$) and anti-symmetric ($\nu_4$ at 1581 cm$^{-1}$) carbon-oxygen stretches as well as the CH deformation mode ($\nu_5$ at 1380 cm$^{-1}$). Formic acid (HCOOH), which is the conjugate acid of the formate ion, is also identified by several bands. The C–O stretch ($\nu_6$) appears at 1239 cm$^{-1}$, while the in-phase C=O stretch ($\nu_2$) is present at 1651 cm$^{-1}$. The $\nu_2$ mode may also occur at 1687 cm$^{-1}$, but this region is dominated by the $\nu_2$ mode of water. The final unambiguous assignment is for formaldehyde ($H_2CO$), which was identified from the C=O stretch at 1726 cm$^{-1}$ ($\nu_2$) and $CH_2$ deformation at 1497 cm$^{-1}$ ($\nu_3$). A very small peak at 1248 cm$^{-1}$ ($\nu_5$) may also belong to formaldehyde. Other ambiguous or unidentified peaks are described in Section 3.4. The spectrum for IRAS 15398-3359 indicates these ices are relatively unprocessed and serve as a baseline upon which other spectra can be compared.

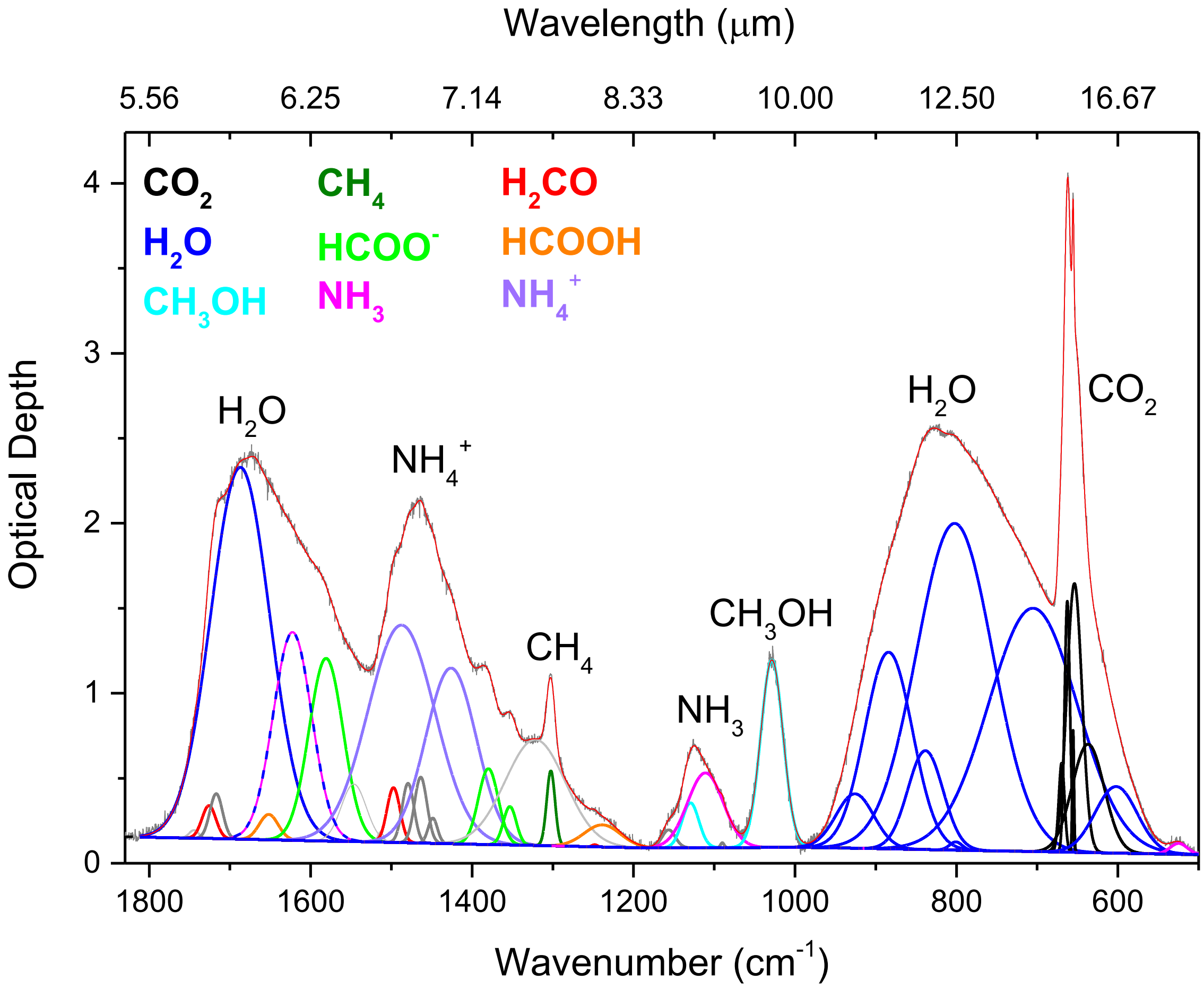

**Figure 4**. Infrared spectrum of IRAS 15398-3359 after continuum removal and peak fitting. The original spectrum (dark gray) is fit with color-coded peaks, which are identified in the legend, that sum to an overall fit (red). Light gray peaks are not assigned to one molecule and are explained in Table 2.

3.2. Ser-emb7

The identifications for Ser-emb7 (Figure 5) are similar to IRAS 15398-3359, but with a few subtle differences. The spectrum shows strong signatures for water ($H_2O$), with broad features peaking at 1685 cm$^{-1}$ ($\nu_2$) and 770 cm$^{-1}$ ($\nu_L$). The double-peaked nature of the strong carbon dioxide band at 660 cm$^{-1}$ ($\nu_2$), which is indicative of pure $CO_2$, is present but less obvious than in the IRAS 15398-3359 spectrum, and the ammonium ($NH_4^+$) cluster peaking at 1465 cm$^{-1}$ ($\nu_4$) is prominent. Relative to IRAS 15398-3359, methanol ($CH_3OH$) shows more pronounced features in Ser-emb7, and the 1030 cm$^{-1}$ ($\nu_8$) band is the second most intense peak observed after carbon dioxide ($\nu_2$). The $\nu_{11}$ peak at 1127 cm$^{-1}$ is also present and more pronounced. Between these peaks is a broad region assigned to ammonia ($NH_3$, 1110 cm$^{-1}$, $\nu_2$). Methane ($CH_4$, 1304 cm$^{-1}$, $\nu_4$) and the formate ion ($HCOO^-$; $\nu_2$ at 1353 cm$^{-1}$, $\nu_5$ at 1382 cm$^{-1}$, and $\nu_4$ at 1581 cm$^{-1}$) are assigned using the same peaks as IRAS 15398-3359, but formic acid (HCOOH) is not observed; this finding suggests an active acid-base chemistry that favors the anionic formate ion. The region between 1200 and 1260 $cm^{-1}$ is devoid of signal except for the small $\nu_5$ peak of formaldehyde ($H_2CO$) at 1247 $cm^{-1}$. The $\nu_2$ in-phase stretch around 1650 $cm^{-1}$ is also absent, and the entire carbonyl (C=O) stretching region is poorly defined. This region likely contains formaldehyde (1717 $cm^{-1}$, $\nu_2$) and other unidentified carbonyl-containing compounds. However, a small new peak at 889 $cm^{-1}$ appears, which we assign to $\nu_{12}$ of ethanol ($CH_3CH_2OH$). The $\nu_{10}$ band may also be present at 1090 $cm^{-1}$, but this peak may have multiple possible assignments. This peak and other uncertain assignments are considered in Section 3.4. While the spectrum of Ser-emb7 is similar to that of IRAS 15398-3359, the presence of ethanol suggests that Ser-emb7 contains mildly processed ices.

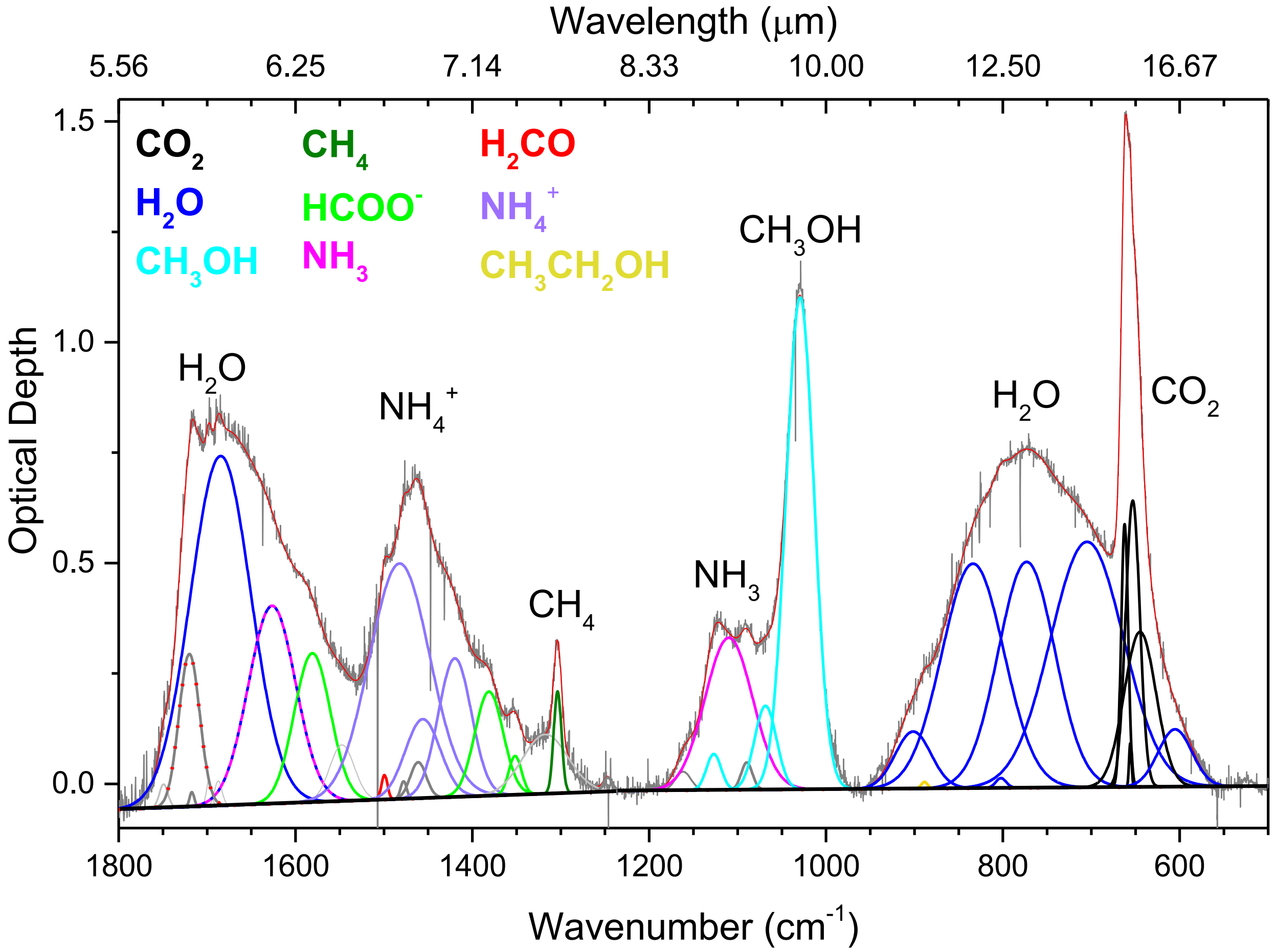

**Figure 5**. Infrared spectrum of Ser-emb7 after continuum removal and peak fitting. The original spectrum (dark gray) is fit with color-coded peaks, which are identified in the legend, that sum to an overall fit (red). Light gray peaks are not assigned to one molecule and are explained in Table 2, and the locally-fitted baseline is black.

3.3. L483

Unlike IRAS 15398-3359 and Ser-emb7, which contain relatively simple compounds with few higher order products, L483 (Figure 6) displays a richer chemistry. Water ($H_2O$; $\nu_2$ at 1690 cm$^{-1}$ and $\nu_L$ at 850 cm$^{-1}$) and carbon dioxide ($CO_2$; $\nu_2$ at 660 cm$^{-1}$) still appear in large quantities but comparatively less than other species, which is opposite to their dominant presence in the other spectra. Pure carbon dioxide is present although its strength is weaker than that seen in IRAS 15398-3359. Relative increases are observed for methanol ($CH_3OH$; $\nu_8$ at 1029 cm$^{-1}$, which is the most intense peak in the spectrum, and $\nu_{11}$ at 1131 cm$^{-1}$), ammonia ($NH_3$; $\nu_2$ at 1109 cm$^{-1}$, $\nu_4$ at 1635 cm$^{-1}$, and $\nu_L$ at 525 cm$^{-1}$), and ammonium ($NH_4^+$, $\nu_{10}$ at 1466 cm$^{-1}$). No significant changes to methane ($CH_4$; $\nu_4$ at 1303 cm$^{-1}$) and formate ($HCOO^-$; $\nu_2$ at 1353 cm$^{-1}$, $\nu_5$ at 1382 cm$^{-1}$, and $\nu_4$ at 1581 cm$^{-1}$) were observed, although formic acid (HCOOH) is identifiable at 1656 cm$^{-1}$ ($\nu_2$, in-phase) along with $\nu_2$ at 1693 cm$^{-1}$, which shares intensity with the $\nu_2$ band of water, and the $\nu_6$ band at 1213 cm$^{-1}$, but this peak may belong to other compounds as well (see Table 2). Formaldehyde ($H_2CO$) is more readily identified with unique bands at 1727 cm$^{-1}$ ($\nu_2$) and 1497 cm$^{-1}$ ($\nu_3$).

In addition to the simple compounds, several larger molecules are observed in L483, and while a few have unique bands, many others share their peaks and thus have ambiguous assignments. Ethanol ($CH_3CH_2OH$) appears more prominently than in Ser-emb7 at 890 cm$^{-1}$ ($\nu_{12}$) and may also contribute to the small but prominent peak at 1090 cm$^{-1}$ ($\nu_{10}$). Two nitrogen-bearing species, methylamine ($CH_3NH_2$) and hydroxylamine ($NH_2OH$), can be possibly identified in L483. A small peak belonging to the deformation mode of the $NH_2$ functional group of methylamine appears at 1642 cm$^{-1}$ ($\nu_4$), while the deformation mode of the $CH_3$ component is seen at 1452 cm$^{-1}$ ($\nu_5$). Two small peaks are also observed for hydroxylamine, both for deformation or rocking of the $NH_2$ functional group, at 1196 cm$^{-1}$ ($\nu_8$) and 1564 cm$^{-1}$ ($\nu_3$). However, the two peaks for each of these nitrogen-bearing compounds are small and limit the detection to a likely assignment until further studies can confirm their presence. Several additional peaks are possible for both methylamine and hydroxylamine but could be assigned to other compounds and thus are considered in the next section. The final peak is a small shoulder of the methane peak at 1308 cm$^{-1}$ that could be one of two notable compounds. It may belong to the sulfur-oxygen stretching mode of sulfur dioxide ($SO_2$, $\nu_3$), which would be the only evidence of sulfur-containing species in these ices. However, it may also belong to a strong overtone of the cyanate ion ($OCN^-$, $2\nu_2$), which would be an indication of cyano/nitrile functional groups from this source. Further ice modeling and gas-phase observations are required to conclude the origin of these absorption peaks. Several additional small peaks were observed in L483 that could have multiple possible assignments, and these are discussed in the next section. Unlike the previous two sources, the COMs observed in L483 indicate that this is a highly processed source.

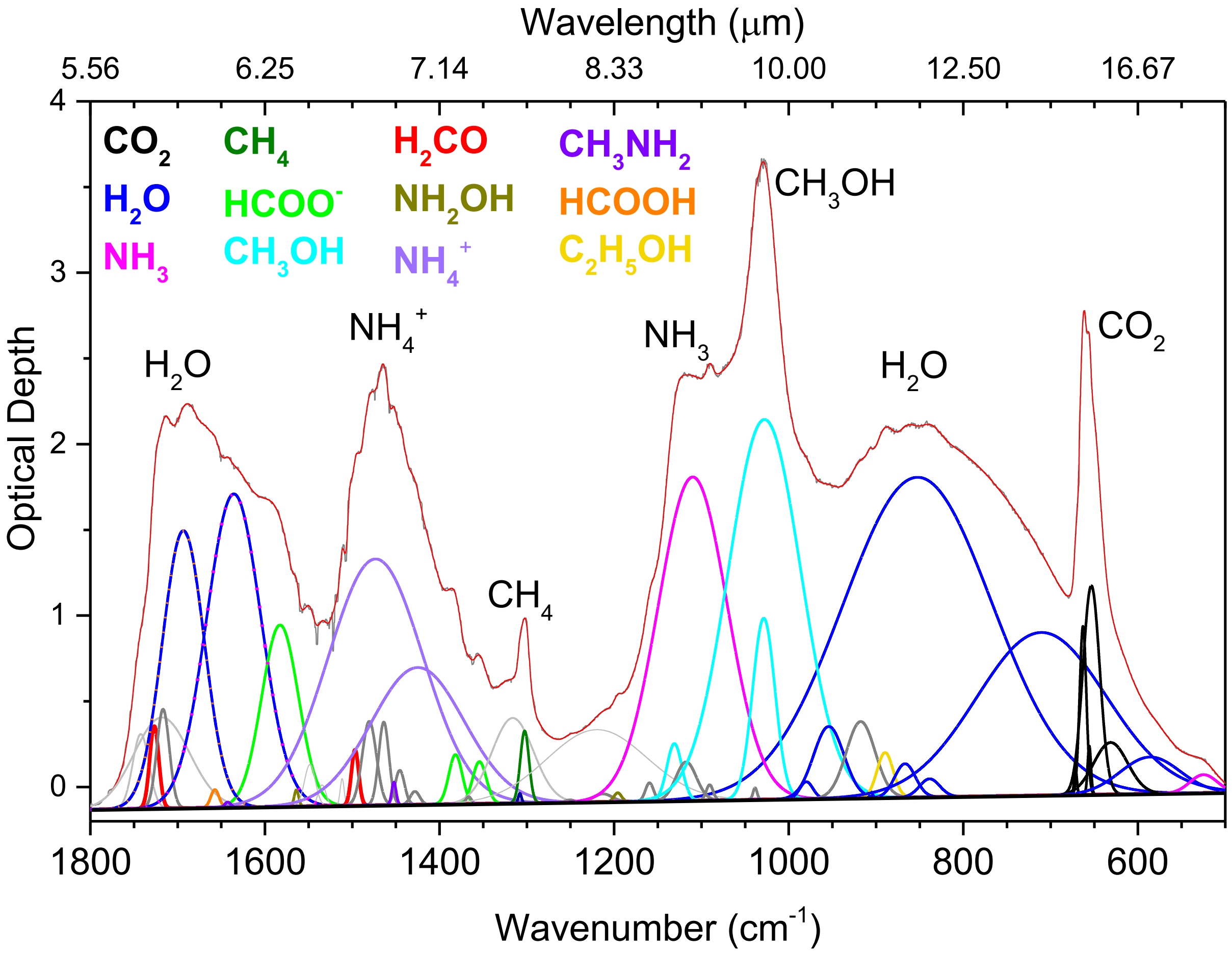

**Figure 6**. Infrared spectrum of L483 after continuum removal and peak fitting. The original spectrum (dark gray) is fit with color-coded peaks, which are identified in the legend, that sum to an overall fit (red). Light gray peaks are not assigned to one molecule and are explained in Table 2.

### 3.4. Ambiguous Assignments

The three spectra contain peaks that allow for confident identification of simple compounds such as water ($H_2O$), carbon dioxide ($CO_2$), methane ($CH_4$), methanol ($CH_3OH$), ammonia/ammonium ($NH_3$/$NH_4^+$), formate/formic acid ($HCOO^-$/ HCOOH), and formaldehyde ($H_2CO$). However, the spectra also display numerous small peaks that may have more than one possible carrier thus providing ambiguous assignments; such ambiguity complicates the identification of more complex molecules as these often have overlapping peaks due to similar functional groups. This is particularly evident for molecules with the carbonyl (C=O) and methyl ($CH_3$) functional group. In all three spectra at 1717 $cm^{-1}$, C=O stretching occurs for molecules such as formic acid (HCOOH, $\nu_2$ out-of-phase), acetaldehyde ($CH_3CHO$, $\nu_4$), methyl formate ($HCOOCH_3$, $\nu_{14}$), acetic acid ($CH_3COOH$, $\nu_4$), and acetone ($CH_3C(O)CH_3$, $\nu_3$). In addition, C=O is observed for various unidentified compounds around 1745 $cm^{-1}$. Small molecules can escape this obscurity, as the carbonyl stretching for formaldehyde occurs at slightly higher energy (1725 $cm^{-1}$, $\nu_2$) and formic acid at slightly lower energy (around 1695 $cm^{-1}$, $\nu_2$).

The methyl-group deformation occurs over a broad range from 1350 to 1500 $cm^{-1}$ and is complicated by overlapping bands from other vibrations. At 1480 $cm^{-1}$, methanol ($CH_3OH$, $\nu_4$) and methylamine ($CH_3NH_2$, $\nu_{12}$) undergo methyl-group deformation, while at the same position, the N–O–H deformation of hydroxylamine ($NH_2OH$, $\nu_4$) occurs. Fortunately, additional transitions improve the likely identification of each of these three compounds, as was described in L483. Methanol also shares a band position with ethane ($C_2H_6$) at 1464 $cm^{-1}$ ($\nu_{10}$ for methanol, $\nu_{11}$ for ethane) as well as with acetone at 1446 $cm^{-1}$ ($\nu_5$ for methanol, $\nu_{21}$ for acetone). This latter band at 1446 $cm^{-1}$ may also result from the intense combination band of acetonitrile ($CH_3CN$, $\nu_7$+$\nu_8$), which if present would be the second possible molecule with a cyano/nitrile functional group after the cyanate ion ($OCN^-$). Furthermore, acetaldehyde ($\nu_5$/$\nu_{10}$) and acetone ($\nu_4$), along with even more complex saturated aldehydes and ketones, contribute to the peak at 1428 $cm^{-1}$ as well as the OH deformation mode of methanol ($\nu_6$). Additional methyl-deformation bands are observed at 1367 $cm^{-1}$ and may involve ethane ($\nu_6$) and acetone ($\nu_{16}$) as well as the C–O deformation of formic acid ($\nu_4$). Also, the deformation vibration for acetaldehyde ($\nu_7$) may contribute to the peak at 1355 $cm^{-1}$, which has been assigned to formate ($\nu_2$). At slightly lower energies, the rocking vibration of the methyl group also occurs and results in overlapping bands. While the small peak observed in each spectrum at 1250 $cm^{-1}$ is assigned to $CH_2$ rocking of formaldehyde ($\nu_5$), this position also relates to $CH_3$ rocking of dimethyl ether ($CH_3OCH_3$, $\nu_5$). This rocking vibration is notably observed at 1160 $cm^{-1}$ for methylamine ($\nu_7$), methyl formate ($\nu_9$), and again for dimethyl ether ($\nu_{20}$). This band could also result from C–O stretching of acetic acid ($\nu_9$). At 1090 $cm^{-1}$, $CH_3$ rocking of ethanol ($CH_3CH_2OH$, $\nu_{10}$) overlaps with the C–O stretching of dimethyl ether ($\nu_{17}$), while at 1039 $cm^{-1}$, $CH_3$ rocking of acetonitrile ($\nu_7$) shares a band position with the C–N stretching mode of methylamine ($\nu_8$).

While vibrations of the methyl group result in a substantial number of overlapping bands, ambiguous assignments involving other functional groups occur as well. At 1213 $cm^{-1}$, C–O stretching of formic acid ($\nu_6$) and methyl formate ($\nu_8$) overlaps with C–C stretching of acetone ($\nu_{17}$), while C–C stretching of acetaldehyde ($\nu_8$) shares a band position with $NH_2$ wagging of hydroxylamine ($\nu_5$) at 1117 $cm^{-1}$. Also, C–O is observed for various unidentified compounds around 1320 $cm^{-1}$. The best example of the difficulty assigning peaks to complex molecules in a mixed ice occurs at 918 $cm^{-1}$, which involves five compounds with different functional groups: N–O stretching of hydroxylamine ($\nu_6$), $NH_2$ wagging of methylamine ($\nu_9$), C–C stretching of acetonitrile ($\nu_4$), C–O stretching of dimethyl ether ($\nu_6$), and O–$CH_3$ stretching of methyl formate ($\nu_{10}$).

In summary, the following molecules might be present, but could *not* be unambiguously identified: ethane ($C_2H_6$), acetonitrile ($CH_3CN$), acetaldehyde ($CH_3CHO$), dimethyl ether ($CH_3OCH_3$), acetic acid ($CH_3COOH$), methyl formate ($HCOOCH_3$), and acetone ($CH_3C(O)CH_3$). Possible formation routes from simple starting materials to these complex organic molecules are presented in the Discussion. Certainly, even larger and more complex molecules could be considered as they would have similar infrared signatures, but for the sake of brevity the simplest homologues were chosen.

**Table 2**. Infrared assignments for each of the three spectra shown in Figures 4-6. Non-unique carriers are given in italics.

| IRAS 15398-3359 Position ($cm^{-1}$) | Ser-emb7 Position ($cm^{-1}$) | L483 Position ($cm^{-1}$) | Assignment | Carrier | Ref. |
|---|---|---|---|---|---|
| 1744 | 1749 | 1742, 1718b | *functional group* | ν(C=O) | 1 |
| 1726 | 1720 | 1727 | $H_2CO$ ($\nu_2$) | ν(C=O) | 2 |
| 1717 | 1717, 1720 | 1717 | HCOOH ($\nu_2$, out-of-phase)<br>*$CH_3CHO$ ($\nu_4$)*<br>*$HCOOCH_3$ ($\nu_{14}$)*<br>*$CH_3COOH$ ($\nu_4$)*<br>*$CH_3C(O)CH_3$ ($\nu_3$)* | ν(C=O) | 3<br>4<br>5, 6<br>7<br>8 |
| 1687 | 1707 | 1693 | $H_2O$ ($\nu_2$)<br>HCOOH ($\nu_2$) | $\delta$(OH)<br>$\nu_s$(C=O) | 9<br>10 |
| 1651 | | 1657 | HCOOH ($\nu_2$, in-phase) | $\nu_s$(C=O) | 3 |
| | | 1642 | $CH_3NH_2$ ($\nu_4$) | $\delta(NH_2)$ | 11 |
| 1623 | 1651 | 1635 | $H_2O$ ($\nu_2$)<br>$NH_3$ ($\nu_4$) | $\delta$(OH)<br>$\delta(NH_3)$ | 9<br>12, 13 |
| 1581 | 1581 | 1583 | $HCOO^-$ ($\nu_4$) | $\nu_a$(CO) | 14 |
| | | 1564 | $NH_2OH$ ($\nu_3$) | $\delta(NH_2)$ | 15, 16 |
| 1497 | 1499 | 1497 | $H_2CO$ ($\nu_3$) | $\delta(CH_2)$ | 2 |
| 1482 | 1478 | 1480 | $CH_3OH$ ($\nu_4$)<br>$CH_3NH_2$ ($\nu_{12}$)<br>$NH_2OH$ ($\nu_4$) | $\delta(CH_3)$<br>$\delta_a(CH_3)$<br>$\delta$(NOH) | 17<br>11<br>15 |
| 1463 | 1461 | 1464 | $CH_3OH$ ($\nu_{10}$)<br>*$C_2H_6$ ($\nu_{11}$)* | $\delta(CH_3)$ | 17<br>9, 18 |
| | | 1452 | $CH_3NH_2$ ($\nu_5$) | $\delta_a(CH_3)$ | 11 |
| 1447 | | 1446 | $CH_3OH$ ($\nu_5$)<br>*$CH_3C(O)CH_3$ ($\nu_{21}$)*<br>*$CH_3CN$ ($\nu_7$+$\nu_8$)* | $\delta_s(CH_3)$<br>$\delta_a(CH_3)$<br>Combination | 17<br>8<br>19, 20 |
| | | 1428 | $CH_3OH$ ($\nu_6$)<br>*$CH_3CHO$ ($\nu_5$,$\nu_{10}$)*<br>*$CH_3C(O)CH_3$ ($\nu_4$)* | $\delta$(OH)<br>$\delta_a(CH_3)$<br>$\delta_a(CH_3)$ | 17<br>4<br>8 |
| 1426, 1487 | 1421, 1452, 1482 | 1424, 1473 | $NH_4^+$ ($\nu_4$) | $\delta$(NH) | 14 |
| 1380 | 1382 | 1382 | $HCOO^-$ ($\nu_5$) | $\delta$(CH) | 14 |
| | | 1367 | HCOOH ($\nu_4$)<br>*$C_2H_6$ ($\nu_6$)*<br>*$CH_3C(O)CH_3$ ($\nu_{16}$)* | $\delta$(CO)<br>$\delta_s(CH_3)$<br>$\delta_s(CH_3)$ | 3<br>9<br>8 |
| 1353 | 1353 | 1355 | $HCOO^-$ ($\nu_2$)<br>*$CH_3CHO$ ($\nu_7$)* | $\nu_s$(CO)<br>$\delta_s(CH_3)$ | 14<br>4 |
| 1324 | 1317 | 1316 | *functional group* | ν(CO) | 1 |
| | | 1308 | *$SO_2$ ($\nu_3$)*<br>*$OCN^-$ ($2\nu_2$)* | ν(SO)<br>Overtone | 21<br>22 |
| 1302 | 1304 | 1303 | $CH_4$ ($\nu_4$) | $\delta$(CH) | 9 |
| 1248 | 1247 | 1251 | $H_2CO$ ($\nu_5$)<br>*$CH_3OCH_3$* ($\nu_5$) | $\rho(CH_2)$<br>$\rho(CH_3)$ | 2<br>4 |
| 1239 | | 1213 | $HCOOH$ ($\nu_6$)<br>*$HCOOCH_3$ ($\nu_8$)*<br>*$CH_3C(O)CH_3$ ($\nu_{17}$)* | $\nu(CO)$<br>$\nu(CO)$<br>$\nu(CCC)$ | 3<br>5<br>8 |
| | | 1196 | $NH_2OH$ ($\nu_8$) | $\rho(NH_2)$ | 15 |
| 1157 | 1161 | 1160 | $CH_3NH_2$ ($\nu_7$)<br>*$CH_3OCH_3$ ($\nu_{20}$)*<br>*$HCOOCH_3$ ($\nu_9$)*<br>*$CH_3COOH$ ($\nu_9$)* | $\rho(CH_3)$<br>$\rho(CH_3)$<br>$\rho(CH_3)$<br>$\nu(CO)$ | 11<br>4<br>5, 6<br>7 |
| 1130 | 1127 | 1131 | $CH_3OH$ ($\nu_{11}$) | $\rho(CH_3)$ | 17 |
| | | 1117 | $NH_2OH$ ($\nu_5$)<br>*$CH_3CHO$ ($\nu_8$)* | $\omega(NH_2)$<br>$\nu(CC)$ | 15<br>4 |
| 1111 | 1110 | 1109 | $NH_3$ ($\nu_2$) | $\delta_s(NH_3)$ | 12, 13 |
| 1090 | 1090 | 1090 | $CH_3CH_2OH$ ($\nu_{10}$)<br>*$CH_3OCH_3$ ($\nu_{17}$)* | $\rho(CH_3)$<br>$\nu_a(CO)$ | 4 |
| | | 1039 | $CH_3NH_2$ ($\nu_8$)<br>*$CH_3CN$ ($\nu_7$)* | $\nu(CN)$<br>$\rho(CH_3)$ | 11<br>19, 20 |
| 1029 | 1030, 1069 | 1029 | $CH_3OH$ ($\nu_8$) | $\nu(CO)$ | 17 |
| 602-927 | 605-901 | 582-979 | $H_2O$ | $\nu_L$ | 9 |
| | | 918 | $NH_2OH$ ($\nu_6$)<br>$CH_3NH_2$ ($\nu_9$)<br>*$CH_3CN$ ($\nu_4$)*<br>*$CH_3OCH_3$ ($\nu_6$)*<br>*$HCOOCH_3$ ($\nu_{10}$)* | $\nu(NO)$<br>$\omega(NH_2)$<br>$\nu(CC)$<br>$\nu_s(CO)$<br>$\nu(CH_3–O)$ | 15<br>11<br>19, 20<br>4<br>5, 6 |
| | 889 | 890 | $CH_3CH_2OH$ ($\nu_{12}$) | $\nu(CC)$ | 4 |
| 637-670 | 646-662 | 631-672 | $CO_2$ ($\nu_2$) | $\delta(CO_2)$ | 9 |
| 525 | 525 | 525 | $NH_3$ | $\nu_L$ | 12 |

**Note.** The "functional group" assignment describes overlapping vibrations for the same functional group among numerous molecules.

**References**. (1) Socrates (2001); (2) Maity et al. (2014); (3) Bisschop et al. (2007); Shimanouchi (1972) (4) van Scheltinga et al. (2018), ; (5) van Scheltinga et al. (2021); (6) Bennett and Kaiser (2007b); (7) Bennett and Kaiser (2007a); (8) Hudson (2018); (9) Turner et al. (2018); (10) Bergantini et al. (2014); (11) Rachid et al. (2021); (12) Zheng and Kaiser (2007a); (13) McClure et al. (2023); (14) Galvez et al. (2010), Maas (1977); (15) Zheng and Kaiser (2010); (16) Yeo and Ford (1990); (17) Marks et al. (2023a); (18) Hudson et al. (2014); (19) Ennis et al. (2017); (20) Rachid et al. (2022); (21) McAnally et al. (2024); (22) Rocha et al. (2024), Rocha et al. (2025)

## 4. Discussion

### 4.1. Reaction Pathways toward Observed COMs

The infrared spectra of the three sources show that the ices consist predominantly of simple molecules in combination with more complex species that likely formed by external irradiation sources such as galactic cosmic rays (GCRs) (Abplanalp et al. 2016b; Bennett et al. 2005; Garrod & Herbst 2006; Garrod et al. 2008; Kaiser 2002; Turner & Kaiser 2020). Laboratory studies that simulate these ices provide valuable information about the formation mechanisms that lead to products through irradiation-initiated chemistry. Starting with the simplest reactants known to exist in interstellar ices, below we present observed neutral-neutral reactions in laboratory ices leading to the formation of complex molecules detected in the JWST spectra (Figure 7). In particular, the focus of this analysis relies on radical-radical recombination pathways as these are the dominant means of production toward complex organic molecules in interstellar ice analogues, although insertion pathways have also been observed (Bennett et al. 2007; Bergner et al. 2017; Förstel et al. 2017; Kaiser & Roessler 1997; Kaiser & Roessler 1998; Turner et al. 2016). These laboratory ices were irradiated with high-energy electrons that simulate the cascade of secondary electrons generated in the wake of GCRs penetrating through the ices (Bennett et al. 2005). Observed products in the spectra, including both confirmed and possible ones based on ambiguous assignments, are considered.

The simple molecules detected in these spectra include water ($H_2O$), carbon dioxide ($CO_2$), methane ($CH_4$), methanol ($CH_3OH$), formaldehyde ($H_2CO$), and ammonia ($NH_3$). Carbon monoxide (CO) is also likely present in these ices and is closely related to the presence of carbon dioxide in irradiated ices, although its fundamental vibration occurs outside the spectral range at 2135 cm$^{-1}$ (Turner et al. 2018) but has been detected, for example, in L483 (Chu et al. 2020) . The initial reaction step upon irradiation involves simple bond cleavage in the reactant molecules accompanied by radical formation:

| | | | |
|---|---|---|---|
| $CH_4$ | → | $CH_3 + H$ | R1 |
| $NH_3$ | → | $NH_2 + H$ | R2 |
| $H_2O$ | → | $OH + H$ | R3 |
| $CH_3OH$ | → | $CH_3O + H$ | R4a |
| $CH_3OH$ | → | $CH_2OH + H$ | R4b |
| $H_2CO$ | → | $HCO + H$ | R5 |

The most common dissociation involves the loss of a hydrogen atom. The loss of hydrogen from methane forms the methyl radical ($CH_3$, Reaction 1, Bennett et al. (2006)), while the amino radical ($NH_2$, Reaction 2) forms from ammonia (Zheng et al. 2008). Water, the most abundant compound in interstellar ices, loses a hydrogen atom to form the hydroxyl radical (OH, Reaction 3, Zheng et al. (2006)), while methanol can lose hydrogen from the carbon or from oxygen producing either the methoxy ($CH_3O$, Reaction 4a) or hydroxymethyl ($CH_2OH$, Reaction 4b) radical (Bennett et al. 2007). The formyl radical (HCO, Reaction 5) results from hydrogen loss of formaldehyde (Bennett et al. 2007). In each case, hydrogen atoms generated by these processes are suprathermal, i.e., they are not in thermal equilibrium with the surrounding ices, and carry kinetic energies up to a few electronvolts (Kaiser et al. 1997; Kaiser & Roessler 1997; Kaiser & Roessler 1998; Morton & Kaiser 2003).

In addition to these radicals, the hydroxycarbonyl (HOCO) radical serves as an important transient species and fundamental molecular building block in interstellar chemistry (Bennett & Kaiser 2007a; Holtom et al. 2005; Ishibashi et al. 2024; Ma et al. 2012; McMurtry et al. 2016; Oba et al. 2010; Turner et al. 2020; Zheng & Kaiser 2007b). This radical can be generated from carbon dioxide via addition of a hydrogen atom (Reaction 6a, Zheng and Kaiser (2007b)) and to a lesser extent from carbon monoxide combining with a hydroxyl radical (Reaction 6b, Bennett et al. (2011)). The barriers to addition can be overcome by the excess kinetic energies of the suprathermal hydrogen atoms (Morton & Kaiser 2003).

$CO_2$ + H → HOCO R6a

CO + OH → HOCO R6b

Barrierless recombination of these radicals can form several of the observed species in the spectra. Formic acid (HCOOH) can result from the recombination of the formyl and hydroxyl radicals (Reaction 7) as well as from hydrogen adding to a HOCO radical (Reaction 8, Bennett et al. (2011)).

HCO + OH → HCOOH R7

HOCO + H → HCOOH R8

The reaction of the formyl radical with methoxyl produces methyl formate ($HCOOCH_3$, Reaction 9, Bennett and Kaiser (2007b)), while the hydroxyl combining with the amino radical generates hydroxylamine ($NH_2OH$, Reaction 10, Tsegaw et al. (2017)).

HCO + $CH_3O$ → $HCOOCH_3$ R9

OH + $NH_2$ → $NH_2OH$ R10

An overview of the considered reactions shows that the methyl radical was involved in a majority of pathways. In the simplest case, two methyl radicals can combine to form ethane ($C_2H_6$, Reaction 11, Bennett et al. (2006)) while a methyl radical may also react with the amino radical to form methylamine ($CH_3NH_2$, Reaction 12, Förstel et al. (2017)), with hydroxymethyl to generate ethanol ($CH_3CH_2OH$, Reaction 13, Bergantini et al. (2018)), with methoxy to produce dimethyl ether ($CH_3OCH_3$, Reaction 14, Bergantini et al. (2017)), and with the formyl radical to create acetaldehyde ($CH_3CHO$, Reaction 15, Bennett et al. (2005)).

| | | | |
|---|---|---|---|
| $CH_3 + CH_3$ | → | $C_2H_6$ | R11 |
| $CH_3 + NH_2$ | → | $CH_3NH_2$ | R12 |
| $CH_3 + CH_2OH$ | → | $CH_3CH_2OH$ | R13 |
| $CH_3 + CH_3O$ | → | $CH_3OCH_3$ | R14 |
| $CH_3 + HCO$ | → | $CH_3CHO$ | R15 |

In addition, the reaction of the HOCO radical from Reaction 6 can combine with a methyl radical to create acetic acid ($CH_3COOH$, Reaction 16, Bennett and Kaiser (2007a)).

| | | | |
|---|---|---|---|
| $CH_3 + HOCO$ | → | $CH_3COOH$ | R16 |

Last, the possible occurrence of acetone ($CH_3C(O)CH_3$), which is the simplest ketone and the only three-carbon compound considered for this analysis, can be explained as a second-generation product. Acetaldehyde, the first-generation product from Reaction 14, can lose a hydrogen atom upon irradiation to generate an acetyl radical ($CH_3CO$, Reaction 17, Singh et al. (2022a)), which when combines with a methyl radical forms acetone (Reaction 18, Singh et al. (2022a)).

| | | | |
|---|---|---|---|
| $CH_3CHO$ | → | $CH_3CO + H$ | R17 |
| $CH_3 + CH_3CO$ | → | $CH_3C(O)CH_3$ | R18 |

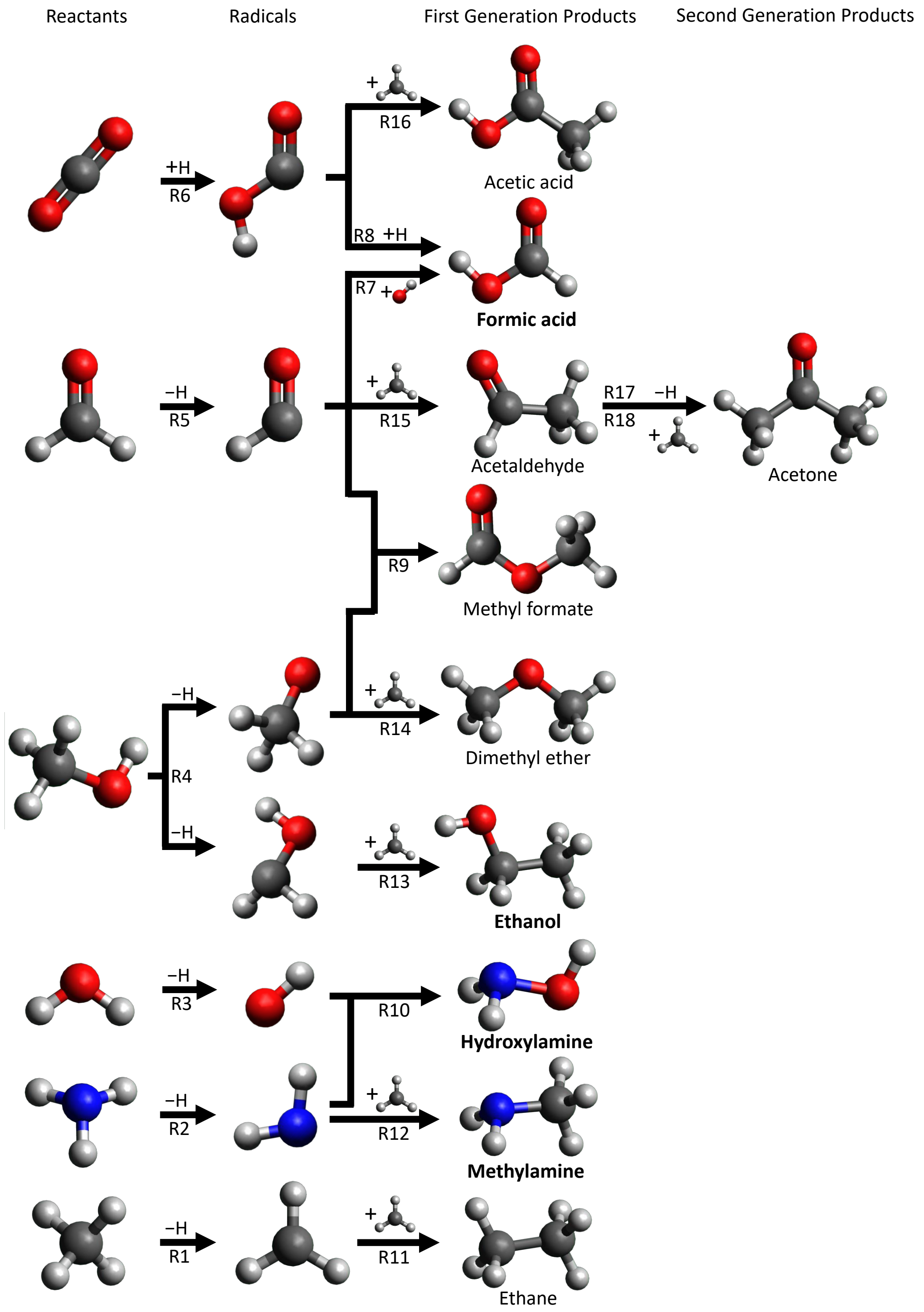

**Figure 7**. Reaction scheme demonstrating the formation of complex organic molecules in the JWST spectra via radical-radical recombination reactions starting from simple reactants such as carbon dioxide ($CO_2$), formaldehyde ($H_2CO$), methanol ($CH_3OH$), water ($H_2O$), ammonia ($NH_3$), and methane ($CH_4$). Each of these pathways have been identified in laboratory ice analogue experiments. Chemical names in bold were identified in the JWST spectra with greater confidence.

## 4.2. Comparing Observed COMs to Laboratory Simulations

While the previous discussion focused on the observed COMs and how laboratory simulation experiments can explain their origins, we now consider if any expected COMs are *missing* from the JWST spectra. The recombination products from the radical species produced by the most abundant reactants, which are identified in Reactions 1–6 and include the methyl ($CH_3$), amino ($NH_2$), hydroxy (OH), hydroxymethyl ($CH_2OH$), methoxy ($CH_3O$), formyl (HCO), and hydroxycarbonyl (HOCO) radicals, are displayed as a grid in Figure 8. The grid is color-coded to indicate COMs that were considered as candidates for absorption features (green) along with those that we have detected in laboratory space simulation experiments but were not assigned in the JWST spectra (blue). The few molecules that belong to neither category are identified in red. Based on these results, the methyl radical was the most prolific at generating COMs as its recombination products with each of the other radicals were potentially observed in the JWST spectra, including ethane ($C_2H_6$, Reaction 11), methylamine ($CH_3NH_2$, Reaction 12), methanol ($CH_3OH$, classified as a reactant), ethanol ($CH_3CH_2OH$, Reaction 13), dimethyl ether ($CH_3OCH_3$, Reaction 14), acetaldehyde ($CH_3CHO$, Reaction 15), and acetic acid ($CH_3COOH$, Reaction 16). Reactions involving the hydroxyl radical with the amino and formyl radicals to form hydroxylamine ($NH_2OH$, Reaction 10) and formic acid (HCOOH, Reaction 7), respectively, were observed, although formic acid may also result from Reaction 8. In addition, methyl formate ($HCOOCH_3$) may form from the combination of the formyl and methoxy radicals (Reaction 9). These results highlight the importance of the methyl radical in COM formation, along with significant contributions from the hydroxyl and formyl radicals.

Three compounds were neither identified in the JWST spectra nor the laboratory experiments: methoxamine ($CH_3ONH_2$), performic acid (HOOCHO), and oxalic acid (HOOCCOOH). Except for oxalic acid, which involves recombination of two HOCO radicals, the proposed mechanism of simple radical-radical recombination toward the formation of the remaining compounds all involve the methoxy ($CH_3O$) radical, which suggests that methoxy may play a more limited role in interstellar chemistry and instead the hydroxymethyl ($CH_2OH$) radical is either the preferred fragmentation pathway of methanol or that the methoxy radical isomerizes to hydroxymethyl (Tachikawa et al. 1994). However, overlapping infrared bands with other ice components could also explain their lack of detection in the JWST spectra. Several compounds not observed in these spectra are highly reactive in terrestrial environments, such as the hydrogen peroxide ($H_2O_2$) and dimethyl peroxide ($CH_3OOCH_3$), although they are stabled in a solid matrix and have been detected laboratory ices (Zheng et al. 2006; Zhu et al. 2019), which similarly describe the simple yet reactive hydrazine ($N_2H_4$) and methanediol ($CH_2(OH)_2$) molecules (Förstel et al. 2015; Zhu et al. 2022). The detections of other products in Figure 8 suffers their *complexity*, and these COMs can be classified as alcohols, aldehydes, or organic acids. The alcohols include methoxymethanol ($CH_3OCH_2OH$, (Zhu et al. 2019)) and ethylene glycol ($HOCH_2CH_2OH$, Zhu et al. (2020)); aldehydes include glycolaldehyde ($OCHCH_2OH$, Maity et al. (2014)) and glyoxal (OCHCHO, Wang et al. (2024)); and acids include carbonic acid ($H_2CO_3$, Zheng and Kaiser (2007b)), glycolic acid ($HOCH_2COOH$, Marks et al. (2023a)), methyl bicarbonate ($CH_3OCOOH$, Marks et al. (2023a)) and glyoxylic acid (OCHCOOH, Eckhardt et al. (2019)). While these compounds may exist in the JWST spectra, their strongest absorption bands overlap with simpler compounds with the same functional groups. In order to observe higher-order COMs in interstellar ices, additional spectra at higher sensitivity may be needed. A fourth category of unobserved molecules in Figure 8 includes nitrogen-carbon-oxygen compounds. In addition to aminomethanol ($NH_2CH_2OH$, Singh et al. (2022b)), carbamic acid ($NH_2COOH$) was detected in non-irradiated ices of $CO_2$ and $NH_3$ and produced by thermal reactions (Marks et al. 2023b), while formamide ($NH_2CHO$, Jones et al. (2011)) in particular was investigated in these JWST ices but no compelling identification was obtained.

Among the 28 predicted compounds in Figure 8, over half were not assigned in our JWST spectra, yet they were synthesized in laboratory ices. However, the majority of these COMs were detected in the *gas phase* of the interstellar medium (ISM) after sublimation of the ices in star forming regions due to the higher sensitivity of rotational spectroscopy compared to FTIR (d'Hendecourt et al. 1982; Tielens et al. 1994; Turner & Kaiser 2020; Vasyunin & Herbst 2013; Westley et al. 1995). In total, six of the predicted compounds were observed in laboratory ices as well as in the gas phase in the ISM, but not detected in the JWST spectra. Formamide ($NH_2CHO$, Rubin et al. (1971)) was identified toward Sgr B2 and was one of the earliest observed molecules in the ISM: before 1971, only 11 other species had been discovered (Woon 2025). The simpler hydrogen peroxide ($H_2O_2$) was detected in 2011 by Bergman et al. (2011), while carbonic acid ($H_2CO_3$) was recently discovered by Sanz-Novo et al. (2023). The three remaining molecules are more complex and each contains two carbon atoms and two oxygen atoms; additionally, Figure 8 predicts each may result from radical-radical reactions involving the hydroxymethyl ($CH_2OH$) radical, which further highlights the potential importance of this radical in interstellar synthesis. Of these three, glycolaldehyde ($OCHCH_2OH$) was the first to be observed (Hollis et al. 2000) followed shortly after by ethylene glycol ($HOCH_2CH_2OH$; Hollis et al. (2002)). Later, one of its isomers, methoxymethanol ($CH_3OCH_2OH$), was also discovered (McGuire et al. 2017). Thus, the addition of the molecules identified in the gas-phase reduces the laboratory-detected molecules without an ISM detection from 15 molecules to 9 molecules, out of a total of 28 molecules predicted in Figure 8.

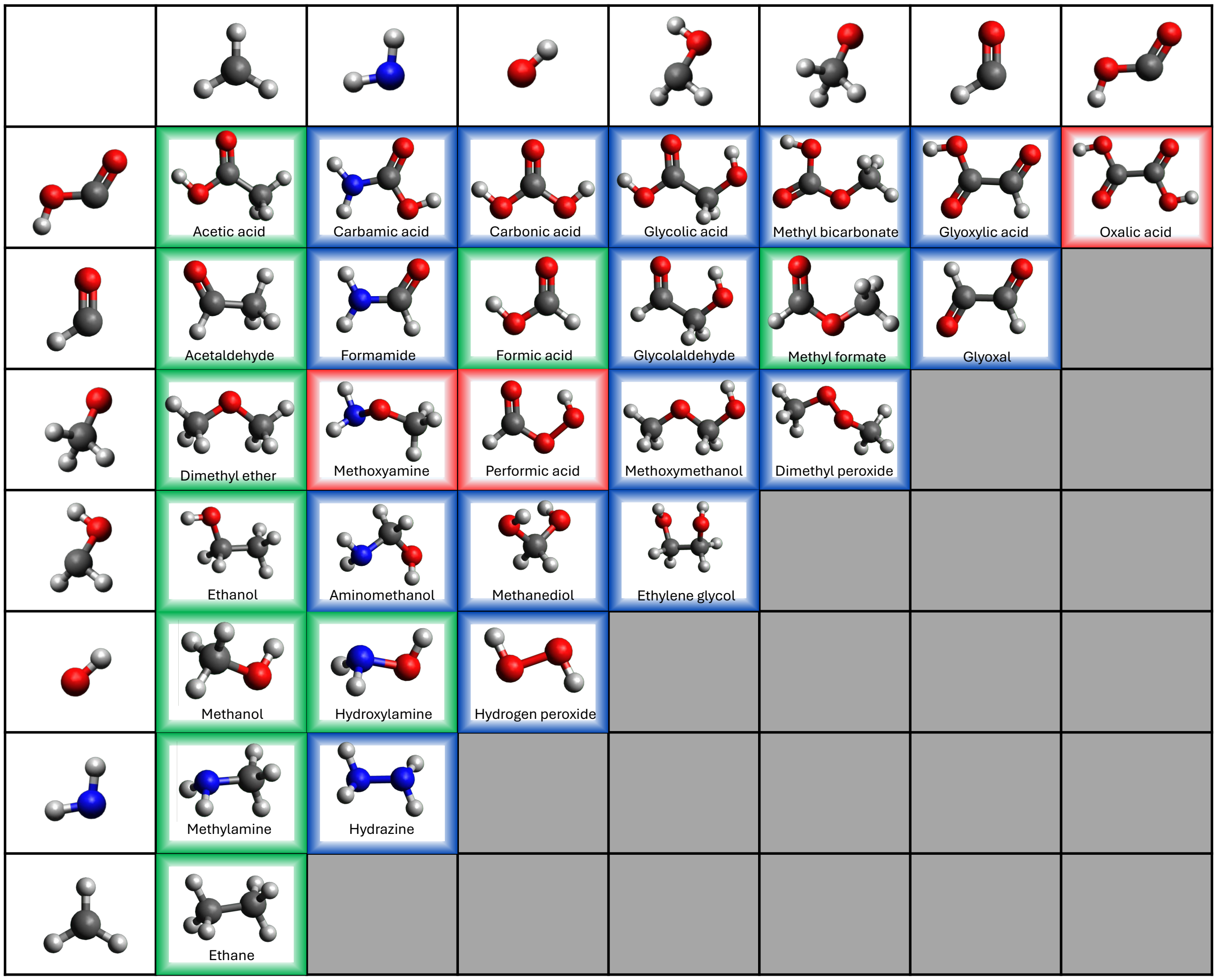

**Figure 8**. The expected products from radical-radical recombination reactions of the most common radicals expected in these ices. Radical reactants are shown on the left and top and include (from left to right, and bottom to top) $CH_3$, $NH_2$, OH, $CH_2OH$, $CH_3O$, HCO, and HOCO, while their products are displayed in the matrix. Products highlighted in green indicate that the product was potentially observed in the JWST spectra (green), while those in blue were not identified in the spectra but were detected in our laboratory simulation experiments. Products in red were not detected by either method.

4.3. Comparing Results to Previous Analyses

While potential signals of COMs were detected in each of the three fully analyzed spectra, caution must be taken before assigning such compounds. COMs should be assigned due to *multiple unique* peaks, as Table 2 demonstrates that most COMs have overlapping vibrational modes. In the present analysis, the best COMs that meet this qualification are formic acid (HCOOH), hydroxylamine ($NH_2OH$), methylamine ($CH_3NH_2$), and ethanol ($CH_3CH_2OH$). However, despite meeting the qualification, some COMs, such as hydroxylamine and methylamine, are more tentative. We attempt to compare the principles in our identification framework to other studies that also investigate COM ices. In McClure et al. (2023), which present the JWST spectra toward two background stars, three COMs were investigated in the $CH_3$ deformation region from 1330–1460 cm$^{-1}$: ethanol ($CH_3CH_2OH$), acetaldehyde ($CH_3CHO$), and acetone ($CH_3C(O)CH_3$), which must be considered as tentative identifications. Several laboratory-based peaks are graphically compared to JWST spectra with $\nu_7$ of acetaldehyde (near 1350 cm$^{-1}$) among the most prominent, but this peak can also be explained by formate ($\nu_2$), and the displayed $\nu_5/\nu_{10}$ band of acetaldehyde near 1430 cm$^{-1}$ does not match the JWST spectra. The most intense band of acetone in this region ($\nu_{16}$ at 1365 cm$^{-1}$) could also be assigned to ethane ($C_2H_6$, $\nu_6$) and formic acid (HCOOH, $\nu_4$), while the three smaller highlighted bands of acetone also overlap other COMs: $\nu_{21}$ at 1440 cm$^{-1}$ with methanol ($\nu_5$) and acetonitrile ($CH_3CN$, $\nu_7$+ $\nu_8$), $\nu_4$ at 1416 cm$^{-1}$ with methanol ($\nu_6$), and $\nu_{17}$ at 1227 cm$^{-1}$ with formic acid ($\nu_6$) and methyl formate ($\nu_8$). Last, the only highlighted ethanol peak (1381 cm$^{-1}$) does not clearly match the JWST spectra and is expected to be weak in a water matrix (van Scheltinga et al. 2018), but also would overlap with formate ($\nu_5$). Similar to the remark made in McClure et al. (2023), detailed modeling together with more sensitive observations are required to confirm the presence of these species.

Furthermore, in Rocha et al. (2024), their analysis of JWST spectra toward two protostars labeled several COMs described as secure detections, including acetaldehyde ($CH_3CHO$), ethanol ($CH_3CH_2OH$), and methyl formate ($HCOOCH_3$), along with acetic acid ($CH_3COOH$) as a tentative detection. In that work, the strategy was to present the most likely ice combination, which includes COMs, determined by algorithmically optimizing the synthetic ice spectra, and their identifications can be compared using the qualifications adopted in our method. First, their table of considered vibrational transitions indicates that these assignments were made using just *one* transition each. The band chosen for ethanol (1383 cm$^{-1}$) is a weak peak, especially in water matrices, and overlaps with the formate ion ($\nu_5$), which was also included among the considered transitions with an absorption coefficient that is three times higher. While both species were modeled to consider the relative contribution in the observed optical depth, the abundance of formate ion would be higher if ethanol is not present at all. Thus, additional confirmation of $\nu_{12}$ at 889 $cm^{-1}$ or $\nu_{10}$ at 1090 $cm^{-1}$ will further support the presence of ethanol. For acetaldehyde, the only peak listed is $\nu_7$ at 1349 $cm^{-1}$, even though McClure et al. (2023) note a second nearby peak that could also be used for acetaldehyde. Additionally, a second peak for the formate ion ($\nu_2$) is listed in the transitions table and would overlap with this $\nu_7$ acetaldehyde peak with an absorption coefficient that is four times higher. In their analysis, a synthetic model that includes formate ion, ethanol, and acetaldehyde is optimized, which might produce a feasible combination to reproduce the observations. However, under our framework, the formate ion was identified based on two matching peaks. Then, the presence of ethanol and acetaldehyde was confirmed using other bands for these two species. For methyl formate, a comparatively strong peak was utilized ($\nu_8$ at 1211 $cm^{-1}$), although this band has similar strength to the overlapping $\nu_2$ of formic acid (HCOOH). The band used to tentatively assign acetic acid was characterized by the authors' laboratory work and was not observed in our JWST spectra. Baseline determination likely introduces significant uncertainty in identifying acetic acid, as demonstrated in Rocha et al. (2024).

### 4.4. Ice Characterization from Spitzer to JWST

JWST improves the sensitivity and spectral resolution of infrared spectroscopy by orders of magnitude, allowing us to reliably decompose complex ice absorption features into individual Gaussian components and associate these features with likely molecular carriers. Figure 9 shows the JWST spectra compared with the Spitzer IRS spectra of the four sources investigated in this study. The Spitzer spectra all have a higher flux density compared to the JWST spectra because of their larger aperture, ~2”-3”. The JWST spectra have notably higher spectral resolution, tracing the detailed shapes of absorption features and enabling the identifications of the features associated with functional groups. In L483 and B335, JWST also provides better sensitivity to detect fainter signals, which were not detected by Spitzer.

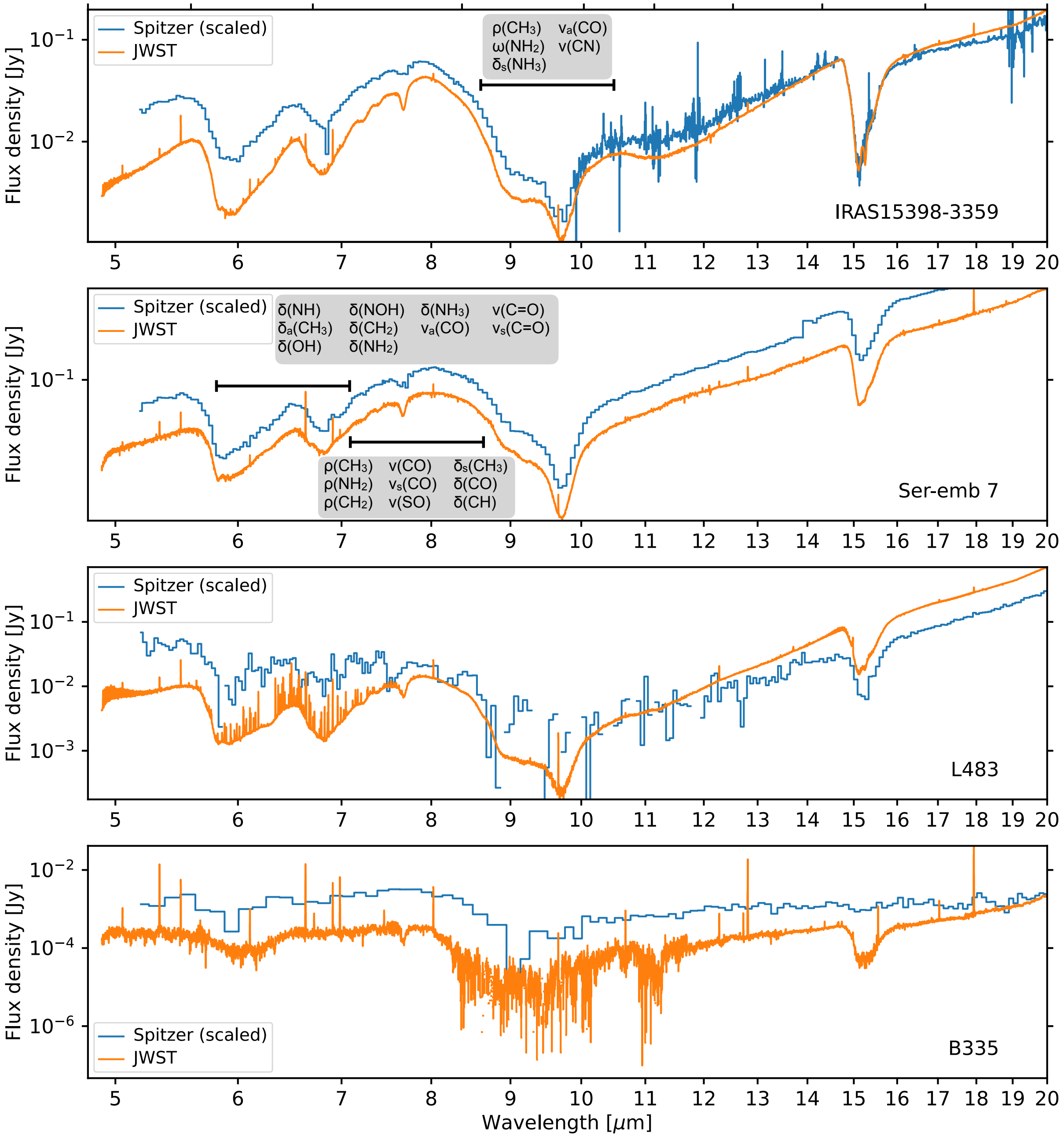

Figure 9. JWST MIRI MRS spectra and the Spitzer IRS spectra of the four CORINOS sources. The Spitzer spectra are scaled down to match the flux density in the JWST spectra for better comparison. The molecular vibrational features distinguishable from the JWST spectra are labeled. The Spitzer spectra are gathered from the Spitzer Heritage Archive. The data were taken in the c2d program (Evans et al. 2009).

## 5. Summary and Outlook

MIRI MRS spectra from JWST were obtained as part of the CORINOS program from four protostellar sources: IRAS 15398-3359, Ser-emb7, L483, and B335. The spectra were processed utilizing continuum spectra and silicate-dust modeling to produce optical depth spectra of ices toward these sources from 4.9–28 μm (2040–360 $cm^{-1}$). A peak-fitting program was applied to assist with spectral identification. These results show that the ices are dominated by simple molecules such as water ($H_2O$), carbon dioxide ($CO_2$), methanol ($CH_3OH$), ammonia/ammonium ($NH_3$/$NH_4^+$), and formic acid/formate ($HCOOH$/$HCOO^-$) with lesser signatures for methane ($CH_4$) and formaldehyde ($H_2CO$). The signals for COMs are weaker, as expected, but detections include hydroxylamine ($NH_2OH$), methylamine ($CH_3NH_2$), and ethanol ($C_2H_5OH$). Other COMs are also likely present and possible examples are displayed in Figure 8, but their overlapping absorption bands hinder the ability to assign these compounds in the ices. The analysis provides as a warning against confidently assigning COMs without sufficient evidence. In order to adequately identify molecules in astronomical ices, a *chemistry-first* approach that incorporates multiple or unique bands is necessary, which works best with a large spectral coverage such as combined NIRSpec and MIRI spectra that probes as many bands as possible. For example, detailed spectra that probes stretching vibrations such as carbon-hydrogen stretching (ν(CH), roughly 3000–3600 $cm^{-1}$) and oxygen-hydrogen stretching (ν(OH), 2800–3100 $cm^{-1}$) (Socrates 2001) would complement the MIRI spectra and provide additional insight into the ambiguous assignments, which often include a $CH_x$ or OH functional group. This cautionary approach is heeded by Pinilla-Alonso et al. (2025), who broadly label portions of spectra as “complex organics” or “aliphatic organics” then discuss the possible functional groups present in the spectra. Furthermore, advances in measurements such as JWST may reveal that as organic molecules become more complex, finding unique peaks will become more challenging (and even impossible) as larger molecules contain functional groups with similar band positions to several other compounds, and while presenting possible assignments is always acceptable, a restrained approach is needed with definitive identifications.

## Acknowledgments

We acknowledge support from the U.S. National Science Foundation (NSF), Division of Astronomical Sciences (AST-2403867), awarded to the University of Hawaii at Manoa (R.I.K). Y.-L.Y. acknowledges support from Grant-in-Aid from the Ministry of Education, Culture, Sports, Science, and Technology of Japan (20H05845 and 25H00676). N. S. acknowledges support from a Grant-in-Aid from the Ministry of Education, Culture, Sports, Science, and Technology of Japan (20H05844 and 20H05845) and a pioneering project in RIKEN (Evolution of Matter in the Universe).

**Competing interests**: The authors declare no competing interests.

## Appendix

### Analysis of B335

The spectra analysis of B335 is limited due to substantial noise level compared to the other three spectra (Figure A1). This is most evident from 950–1250 $cm^{-1}$, which was excluded from the peak-fitting program. The broad yet prominent features for water ($\nu_L$ at 750–850 $cm^{-1}$ and $\nu_2$ at 1600-1700 $cm^{-1}$) are present, with the latter likely having contributions from ammonia ($NH_3$, $\nu_4$). The strong carbon dioxide ($CO_2$) peak is clearly visible ($\nu_2$ at 660 $cm^{-1}$) along with the smaller methane ($CH_4$) absorption at 1302 $cm^{-1}$ ($\nu_4$). The noise prevents a clear assignment of methanol ($CH_3OH$) at 1030 $cm^{-1}$ ($\nu_8$), and the nearby band of ammonia ($\nu_2$ at 1110 $cm^{-1}$) is also obscured. The region from 1400–1500 $cm^{-1}$ that contains prominent ammonium ($NH_4^+$) peaks in the other spectra is much weaker in B335, and thus the presence of ammonium is uncertain. However, the formate counterion may be present at 1355 $cm^{-1}$ ($\nu_2$) and 1382 $cm^{-1}$ ($\nu_5$). While B335 shows similarities to the other three spectra, the noise level prevents the identification of some commonly observed compounds and any possible COMs that may be present.

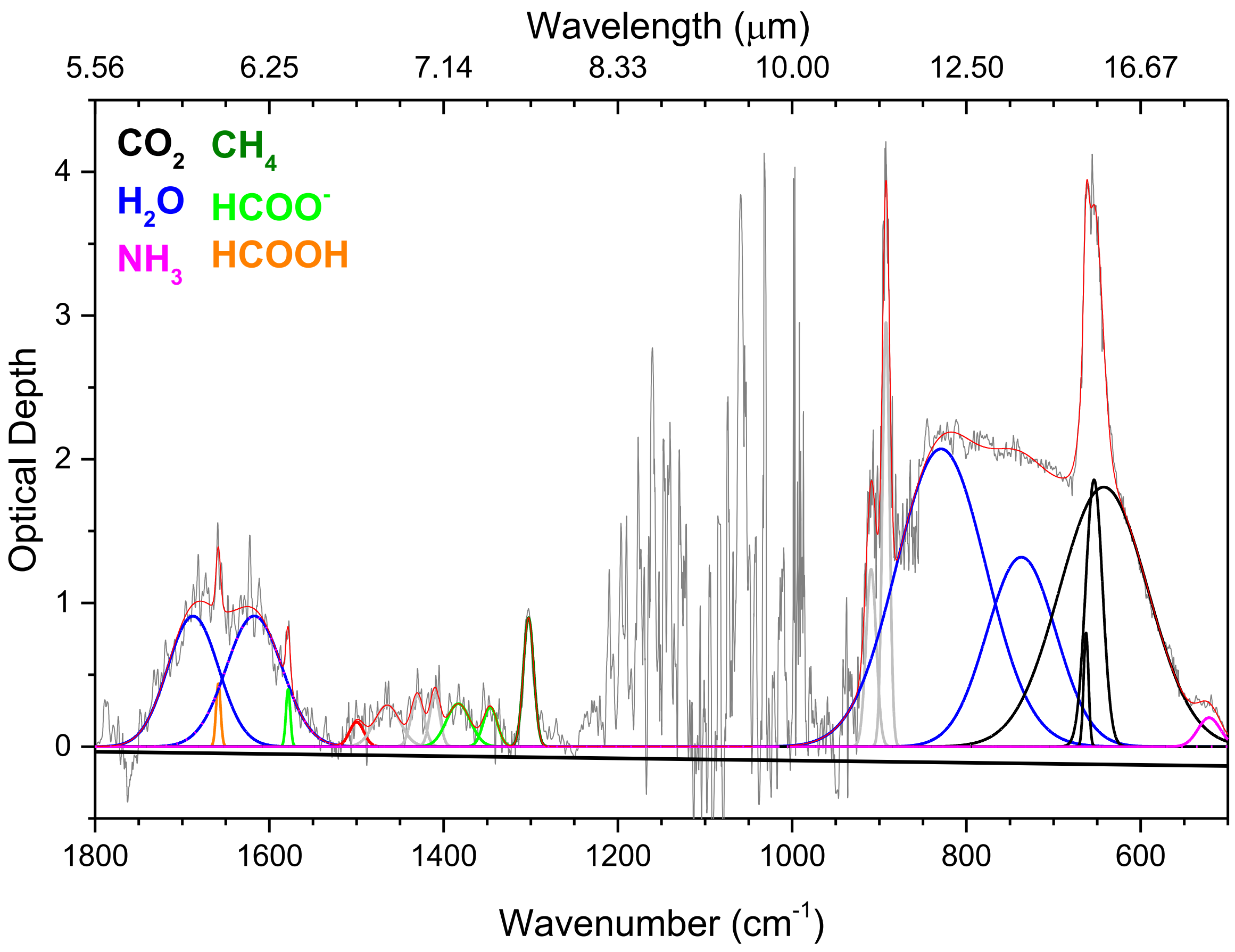

**Figure A1**. Infrared spectrum of B335 after continuum removal and peak fitting. The original spectrum (dark gray) is fit with color-coded peaks, which are identified in the legend, that sum to an overall fit (red). Light gray peaks are not assigned.

**References**

Abplanalp, M. J., Förstel, M., & Kaiser, R. I. 2016a, CPL, 644, 79
Abplanalp, M. J., Gozem, S., Krylov, A. I., et al. 2016b, PNAS, 113, 7727
Aikawa, Y., Kamuro, D., Sakon, I., et al. 2012, A&A, 538, A57
Belloche, A., Maury, A., Maret, S., et al. 2020, A&A, 635, A198
Bennett, C. J., Chen, S. H., Sun, B. J., Chang, A. H. H., & Kaiser, R. I. 2007, ApJ, 660, 1588
Bennett, C. J., Hama, T., Kim, Y. S., Kawasaki, M., & Kaiser, R. I. 2011, ApJ, 727, 27
Bennett, C. J., Jamieson, C. S., Osamura, Y., & Kaiser, R. I. 2005, ApJ, 624, 1097
Bennett, C. J., Jamieson, C. S., Osamura, Y., & Kaiser, R. I. 2006, ApJ, 653, 792
Bennett, C. J., & Kaiser, R. I. 2007a, ApJ, 660, 1289
Bennett, C. J., & Kaiser, R. I. 2007b, ApJ, 661, 899
Bergantini, A., Góbi, S., Abplanalp, M. J., & Kaiser, R. I. 2018, ApJ, 852, 70
Bergantini, A., Maksyutenko, P., & Kaiser, R. I. 2017, ApJ, 841, 96
Bergantini, A., Pilling, S., Rothard, H., Boduch, P., & Andrade, D. 2014, MNRAS, 437, 2720
Bergman, P., Parise, B., Liseau, R., et al. 2011, A&A, 531, L8
Bergner, J. B., Martín-Doménech, R., Öberg, K. I., et al. 2019, ESC, 3, 1564
Bergner, J. B., Öberg, K. I., & Rajappan, M. 2017, ApJ, 845, 29
Bergner, J. B., Öberg, K. I., Rajappan, M., & Fayolle, E. C. 2016, ApJ, 829, 85
Bisschop, S., Fuchs, G., Boogert, A., Van Dishoeck, E., & Linnartz, H. 2007, A&A, 470, 749
Boogert, A. A., Gerakines, P. A., & Whittet, D. C. 2015, ARA&A, 53
Boogert, A. C., Pontoppidan, K. M., Knez, C., et al. 2008, ApJ, 678, 985
Booth, A. S., Walsh, C., Terwisscha van Scheltinga, J., et al. 2021, NatAs, 5, 684
Bottinelli, S., Boogert, A. A., Bouwman, J., et al. 2010, ApJ, 718, 1100
Bradley, L., Sipőcz, B., Robitaille, T., et al. 2024, Zenodo, astropy/photutils: 1.12.0
Brunken, N., Boogert, A., van Dishoeck, E., et al. 2025, ESC, 9, 1992
Brunken, N. G., Rocha, W. R., Van Dishoeck, E. F., et al. 2024, A&A, 685, A27
Bushouse, H., Eisenhamer, J., Dencheva, N., et al. 2024, Zenodo, JWST Calibration Pipeline (1.15.1)
Chen, Y., Rocha, W. R. M., van Dishoeck, E. F., et al. 2024, A&A, 690, A205
Chu, L. E., Hodapp, K., & Boogert, A. 2020, ApJ, 904, 86
Codella, C., Ceccarelli, C., Caselli, P., et al. 2017, A&A, 605, L3
d'Hendecourt, L., Allamandola, L., Baas, F., & Greenberg, J. 1982, A&A, 109, L12
Dominik, C., Min, M., & Tazaki, R. 2021, Astrophysics Source Code Library, ascl: 2104.010
Dorschner, J., Begemann, B., Henning, T., Jaeger, C., & Mutschke, H. 1995, A&A, 300, 503
Eckhardt, A. K., Bergantini, A., Singh, S. K., Schreiner, P. R., & Kaiser, R. I. 2019, ACIE, 58, 5663
Ehrenfreund, P., Boogert, A. C. A., Gerakines, P. A., Tielens, A. G. G. M., & van Dishoeck, E. F. 1997, A&A, 328, 649
Ennis, C., Auchettl, R., Ruzi, M., & Robertson, E. 2017, PCCP, 19, 2915
Enoch, M. L., Evans, N. J., Sargent, A. I., & Glenn, J. 2009, ApJ, 692, 973
Evans, N. J., Dunham, M. M., Jorgensen, J. K., et al. 2009 ApJS, 181, 321
Evans, N. J., Yang, Y.-L., Green, J. D., et al. 2023, ApJ, 943, 90
Federman, S. A., Megeath, S. T., Rubinstein, A. E., et al. 2024, ApJ, 966, 41
Fleming, I., & Williams, D. 2019, in Spectroscopic Methods in Organic Chemistry, Cham: Springer International Publishing, 85
Förstel, M., Bergantini, A., Maksyutenko, P., Góbi, S., & Kaiser, R. I. 2017, ApJ, 845, 83
Förstel, M., Maksyutenko, P., Jones, B. M., et al. 2015, ChemPhysChem, 16, 3139
Galvez, O., Maté, B., Herrero, V. J., & Escribano, R. 2010, ApJ, 724, 539
Gardner, J. P., Mather, J. C., Abbott, R., et al. 2023, PASP, 135, 068001
Garrod, R. T., & Herbst, E. 2006, A&A, 457, 927
Garrod, R. T., Weaver, S. L. W., & Herbst, E. 2008, ApJ, 682, 283

Garrod, R. T., & Widicus Weaver, S. L. 2013, ChRv, 113, 8939
Gibb, E. L., Whittet, D. C. B., Boogert, A. C. A., & Tielens, A. G. G. M. 2004, ApJS, 151, 35
GRAMS/AI-Spectroscopy-Software: Thermo Scientific 2002
Hollis, J. M., Lovas, F. J., & Jewell, P. R. 2000, ApJ, 540, L107
Hollis, J. M., Lovas, F. J., Jewell, P. R., & Coudert, L. 2002, ApJ, 571, L59
Holtom, P., Bennett, C., Osamura, Y., Mason, N., & Kaiser, R. 2005, ApJ, 626, 940
Hudson, R. 2018, PCCP, 20, 5389
Hudson, R., Gerakines, P., & Moore, M. 2014, Icarus, 243, 148
Hudson, R. L., & Gerakines, P. A. 2018, ApJ, 867, 138
Imai, M., Sakai, N., Oya, Y., et al. 2016, ApJL, 830, L37
Ishibashi, A., Molpeceres, G., Hidaka, H., et al. 2024, ApJ, 976, 162
Jacobsen, S. K., Jørgensen, J. K., Di Francesco, J., et al. 2019, A&A, 629, A29
Jones, B. M., Bennett, C. J., & Kaiser, R. I. 2011, ApJ, 734, 78
Jørgensen, J. K., Belloche, A., & Garrod, R. T. 2020, ARA&A, 58, 727
Jørgensen, J. K., Visser, R., Sakai, N., et al. 2013, ApJL, 779, L22
Kaiser, R., Eich, G., Gabrysch, A., & Roessler, K. 1997, ApJ, 484, 487
Kaiser, R., & Roessler, K. 1997, ApJ, 475, 144
Kaiser, R. I. 2002, ChRv, 102, 1309
Kaiser, R. I., & Roessler, K. 1998, ApJ, 503, 959
Kemper, F., Vriend, W., & Tielens, A. 2004, ApJ, 609, 826
Kim, J., Lee, J.-E., Jeong, W.-S., et al. 2022, ApJ, 935, 137
Law, D. R., Morrison, J. E., Argyriou, I., et al. 2023, AJ, 166, 45
López-Sepulcre, A., Codella, C., Ceccarelli, C., Podio, L., & Robuschi, J. 2024, A&A, 692, A120
Ma, J., Li, J., & Guo, H. 2012, JPCL, 3, 2482
Maas, J. 1977, AcSpA, 33, 761
Maity, S., Kaiser, R. I., & Jones, B. M. 2014, FaDi, 168, 485
Marks, J. H., Wang, J., Evseev, M. M., et al. 2023a, ApJ, 942, 43
Marks, J. H., Wang, J., Sun, B.-J., et al. 2023b, ACSCS, 9, 2241
McAnally, M., Bocková, J., Herath, A., et al. 2024, NatCo, 15, 4409
McClure, M. K., Rocha, W. R. M., Pontoppidan, K. M., et al. 2023, NatAs, 7, 431
McGuire, B. A., Shingledecker, C. N., Willis, E. R., et al. 2017, ApJL, 851, L46
McMurtry, B. M., Turner, A. M., Saito, S. E., & Kaiser, R. I. 2016, CP, 472, 173
Morton, R. J., & Kaiser, R. I. 2003, P&SS, 51, 365
Nazari, P., Rocha, W., Rubinstein, A., et al. 2024, A&A, 686, A71
Oba, Y., Watanabe, N., Kouchi, A., Hama, T., & Pirronello, V. 2010, ApJL, 712, L174
Öberg, K. I., Boogert, A. A., Pontoppidan, K. M., et al. 2008, ApJ, 678, 1032
Ohashi, N., Tobin, J. J., Jørgensen, J. K., et al. 2023, ApJ, 951, 8
Okoda, Y., Oya, Y., Francis, L., et al. 2023, ApJ, 948, 127
Okoda, Y., Oya, Y., Imai, M., et al. 2022, ApJ, 935, 136
Okoda, Y., Oya, Y., Sakai, N., Watanabe, Y., & Yamamoto, S. 2020, ApJ, 900, 40
Pinilla-Alonso, N., Brunetto, R., De Prá, M. N., et al. 2025, NatAs, 9, 230
Pontoppidan, Klaus M., Boogert, A. C. A., Fraser, Helen J., et al. 2008, ApJ, 678, 1005
Rachid, M., Brunken, N., De Boe, D., et al. 2021, A&A, 653, A116
Rachid, M., van Scheltinga, J. T., Koletzki, D., & Linnartz, H. 2020, A&A, 639, A4
Rachid, M. G., Rocha, W., & Linnartz, H. 2022, A&A, 665, A89
Rayalacheruvu, P., Majumdar, L., Rocha, W., et al. 2025, doi: 10.48550/arXiv.2506.15358
Rieke, G. H., Wright, G. S., Böker, T., et al. 2015, PASP, 127, 584
Rocha, W., Rachid, M., Olsthoorn, B., et al. 2022, A&A, 668, A63
Rocha, W. R., Perotti, G., Kristensen, L. E., & Jørgensen, J. K. 2021, A&A, 654, A158
Rocha, W. R. M., McClure, M. K., Sturm, J. A., et al. 2025, A&A, 693, A288

Rocha, W. R. M., van Dishoeck, E. F., Ressler, M. E., et al. 2024, A&A, 683, A124
Rubin, R., Swenson Jr, G., Benson, R., Tigelaar, H., & Flygare, W. 1971, ApJ, 169, L39
Santos, J. C., Enrique-Romero, J., Lamberts, T., Linnartz, H., & Chuang, K.-J. 2024, ESC, 8, 1646
Sanz-Novo, M., Rivilla, V. M., Jiménez-Serra, I., et al. 2023, ApJ, 954, 3
Schutte, W., Boogert, A., Tielens, A., et al. 1999, A&A, 343, 966
Schutte, W., & Khanna, R. 2003, A&A, 398, 1049
Scibelli, S., Shirley, Y., Megías, A., & Jiménez-Serra, I. 2024, MNRAS, 533, 4104
Shimanouchi, T. 1972, Tables of Molecular Vibrational Frequencies Consolidated Volume I, National Bureau of Standards, US Government Printing Office Washington, DC
Shimanouchi, T. 1973, Tables of Molecular Vibrational Frequencies, National Bureau of Standards, US Government Printing Office Washington, DC
Singh, S. K., Kleimeier, N. F., Eckhardt, A. K., & Kaiser, R. I. 2022a, ApJ, 941, 103
Singh, S. K., Zhu, C., La Jeunesse, J., Fortenberry, R. C., & Kaiser, R. I. 2022b, NatCo, 13, 375
Slavicinska, K., Boogert, A., van Dishoeck, E., et al. 2025, A&A, 693, A146
Slavicinska, K., Rachid, M. G., Rocha, W. R. M., et al. 2023, A&A, 677, A13
Socrates, G. 2001, Infrared and Raman Characteristic Group Frequencies, Third Edition (New York: John Wiley & Sons, Ltd.)
Tachikawa, H., Lunell, S., Törnkvist, C., & Lund, A. 1994, JMoSt: THEOCHEM, 304, 25
Tielens, A., McKee, C., Seab, C., & Hollenbach, D. 1994, ApJ, 321
Tsegaw, Y. A., Góbi, S., Förstel, M., et al. 2017, JPCA, 121, 7477
Turner, A. M., Abplanalp, M. J., Blair, T. J., Dayuha, R., & Kaiser, R. I. 2018, ApJS, 234, 6
Turner, A. M., Abplanalp, M. J., & Kaiser, R. I. 2016, ApJ, 819, 97
Turner, A. M., & Kaiser, R. I. 2020, Acc. Chem. Res., 53, 2791
Turner, A. M., Koutsogiannis, A. S., Kleimeier, N. F., et al. 2020, ApJ, 896, 88
Tyagi, H., Manoj, P., Narang, M., et al. 2025, ApJ, 983, 110
van Scheltinga, J. T., Ligterink, N., Boogert, A., Van Dishoeck, E., & Linnartz, H. 2018, A&A, 611, A35
van Scheltinga, J. T., Marcandalli, G., McClure, M. K., Hogerheijde, M. R., & Linnartz, H. 2021, A&A, 651, A95
Vasyunin, A., & Herbst, E. 2013, ApJ, 769, 34
Wang, J., Turner, A. M., Marks, J. H., et al. 2024, ApJ, 967, 79
Westley, M., Baragiola, R., Johnson, R., & Baratta, G. 1995, Natur, 373, 405
Woon, D. E. Sept 2025, The Astrochymist, astrochymist.org
Wright, G. S., Rieke, G. H., Glasse, A., et al. 2023, PASP, 135, 048003
Yang, Y.-L., Green, J. D., Pontoppidan, K. M., et al. 2022, ApJL, 941, L13
Yang, Y.-L., Sakai, N., Zhang, Y., et al. 2021, ApJ, 910, 20
Yeo, G., & Ford, T. 1990, JMoSt, 217, 307
Zhang, C., Leyva, V., Wang, J., et al. 2024, PNAS, 121, e2320215121
Zhang, C., Wang, J., Turner, A. M., Young, L. A., & Kaiser, R. I. 2025a, ApJS, 278, 30
Zhang, C., Wang, J., Turner, A. M., Young, L. A., & Kaiser, R. I. 2025b, ApJS, 279, 1
Zhang, C., Young, L. A., & Kaiser, R. I. 2025c, ApJ, 980, 248
Zheng, W., Jewitt, D., & Kaiser, R. I. 2006, ApJ, 639, 534
Zheng, W., Jewitt, D., Osamura, Y., & Kaiser, R. I. 2008, ApJ, 674, 1242
Zheng, W., & Kaiser, R. I. 2007a, CPL, 440, 229
Zheng, W., & Kaiser, R. I. 2007b, CPL, 450, 55
Zheng, W., & Kaiser, R. I. 2010, JPCA, 114, 5251
Zhu, C., Frigge, R., Bergantini, A., Fortenberry, R. C., & Kaiser, R. I. 2019, ApJ, 881, 156
Zhu, C., Kleimeier, N. F., Turner, A. M., et al. 2022, PNAS, 119, e2111938119
Zhu, C., Turner, A. M., Meinert, C., & Kaiser, R. I. 2020, ApJ, 889, 134